\documentclass[12pt]{article}

\usepackage{color}

\usepackage{amsmath}

\usepackage{graphicx}
\usepackage{float}
\usepackage{amsmath}
\usepackage{amsthm}
\usepackage{amssymb}
\usepackage{amscd}
\usepackage{amsfonts}
\usepackage{makeidx}
\usepackage{enumerate}      
\usepackage{IEEEtrantools}
\usepackage{mathrsfs}

\usepackage{amsmath,empheq}

\usepackage[latin1]{inputenc}

\usepackage [all]{xy}
\usepackage{cases}

\usepackage{amsmath}

\usepackage{cancel}

\usepackage{mathrsfs}

\newtheorem{teo}{Theorem}[section]
    \newtheorem{lem}[teo]{Lemma}
    \newtheorem{prop}[teo]{Proposition}

    \newtheorem{obs}[teo]{Remark}
    \newtheorem{exa}[teo]{Example}

    \newtheorem*{dem}{\textsc{Proof}}
    \newcommand{\bdem}{\begin{dem}}
    \newcommand{\edem}{\end{dem}}
     \newcommand{\be}{\begin{equation}}
    \newcommand{\ee}{\end{equation}}
     \newcommand{\ba}{\begin{array}}
    \newcommand{\ea}{\end{array}}
\newcommand{\beqn}{\begin{eqnarray}}
    \newcommand{\eeqn}{\end{eqnarray}}
    \newcommand{\bl}{\begin{lem}}
    \newcommand{\el}{\end{lem}}
    \newcommand{\bp}{\begin{prop}}
    \newcommand{\ep}{\end{prop}}
\newcommand{\ds}{\displaystyle}

    \newcommand{\R}{\mathbb{R}}
    \newcommand{\C}{\mathbb{C}}
     
     \newcommand{\no}{\noindent}
     \newcommand{\Res}{{\rm Res}}

   \providecommand{\abs}[1]{\lvert#1\rvert}
\providecommand{\norm}[1]{\lVert#1\rVert}

    \IEEEyessubnumber

\def\Re {{\rm Re\, }}                                       
 \def\Im {{\rm Im\,}}                                       
\def\Res {{\rm Res\,}}  

\begin{document}
\title{Discrete and embedded trapped modes in a plane quantum waveguide with a small obstacle: exact solutions}

\author{
\large{ P.  Zhevandrov$^1$},\\
\large{A.  Merzon$^2$},\\
\large{M.I. Romero Rodr\'iguez$^3$}\\
\large{and J.E. De la Paz M\'endez$^4$}\\
{\small{\it $^1$ Facultad de  Ciencias F\'\i sico-Matem\'aticas, Universidad Michoac{a}na}},\\[-2mm]
{\small  Morelia, Michoac\'{a}n, M\'{e}xico}\\[-2mm]
{\small{\it $^2$ Instituto de F\'\i sica y  Matem\'aticas, Universidad Michoac{a}na}},\\[-2mm]
{\small  Morelia, Michoac\'{a}n, M\'{e}xico}\\[-2mm]
{\small{\it $^3$ Facultad de Ciencias B\'asicas y Aplicadas, Universidad Militar Nueva Granada}},\\[-2mm]
{\small{Bogot\'a, Colombia}},\\[-2mm]
{\small{\it $^4$ Facultad de Matem\'aticas II, Universidad Aut\'onoma de Guerrero}},\\[-2mm]
{\small{Cd. Altamirano, Guerrero, M\'exico}}\\[-2mm]
{\small{\it E-mails}: pzhevand@umich.mx,}
{\small anatoli.merzon@umich.mx,}\\[-2mm] {\small maria.romeror@unimilitar.edu.co,} {\small jeligio12@gmail.com}}
\maketitle


\begin{abstract}
  
Exact solutions describing trapped modes in a plane quantum waveguide with a small rigid obstacle are constructed in the form of convergent series in powers of the small parameter characterizing the smallness of the obstacle. The terms of this series are expressed through the solution of the exterior Neumann problem for the Laplace equation describing the flow of unbounded fluid past the inflated obstacle.\\
The exact solutions obtained describe discrete eigenvalues of the problem under certain geometric conditions, and, when the obstacle is symmetric, these solutions describe embedded eigenvalues. For obstacles symmetric with respect to the centerline of the waveguide, the existence of embedded trapped modes is known (due to the decomposition trick of the domain of the corresponding differential operator) even without the smallness assumption. We construct these solutions in an explicit form for small obstacles. For obstacles symmetric with respect to the vertical axis, we find embedded trapped modes for a specific vertical displacement of the obstacle.     
\end{abstract}

\maketitle

\section{Introduction}
\setcounter{equation}{0}

\begin{figure}[htbp]
\centering
\includegraphics[scale=0.27]{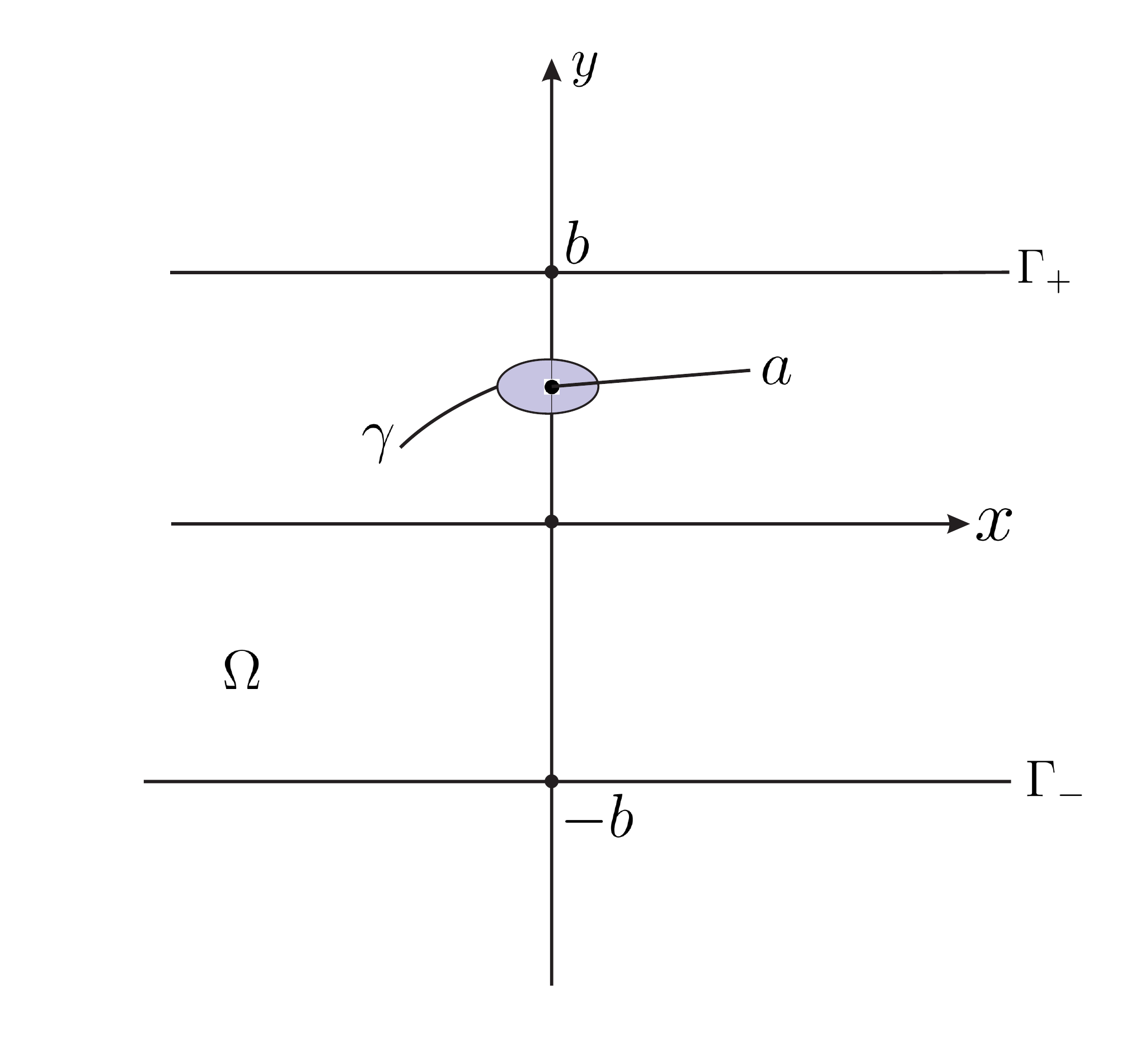}
\caption{Small obstacle in a waveguide}\label{Gam_d}
\end{figure}

The present paper is devoted to the study of trapped modes in a plane quantum waveguide with a small rigid obstacle. Mathematically, this reduces to the study of the following boundary value problem (see Fig.\;\ref{Gam_d}):
\begin{align}\label{PD1}
&-\Delta u=k^2 u,\quad x,y\in\Omega,
\\\nonumber\\\label{PD2}
&u\Big\vert_{\Gamma_{\pm}}=0,\quad \ds\frac{\partial u}{\partial n}\Bigg\vert_{\gamma}=0,
\end{align}
where $k^2$ is a spectral parameter, $\gamma=\Big\lbrace x=\varepsilon X(t),~y=a+\varepsilon Y(t)\Big\rbrace$ is the boundary of the obstacle and $\Omega$ is the interior of the strip $\lbrace-b<y<b\rbrace$ without the obstacle (i.e., the exterior of $\gamma$), $\Gamma_{\pm}=\Big\lbrace x\in\R,~y=\pm b\Big\rbrace$.
We assume that $\abs{a}<b$, $X(t)$ and $Y(t)$ are $2\pi$-periodic,  $C^{\infty}$-functions with zero mean, $\ds\int\limits_{-\pi}^{\pi} X(t)\;dt=\ds\int\limits_{-\pi}^{\pi} Y(t)\;dt=0$; see Figure \ref{Gam_d}. Trapped modes are, by definition, nontrivial solutions of (\ref{PD1}), (\ref{PD2}) which decay at infinity (i.e. as $\abs{x}\to\infty$). The corresponding values of $k^2$ are eigenvalues of (\ref{PD1}), (\ref{PD2}). 

It is well-known that the spectrum of (\ref{PD1}), (\ref{PD2}) without the obstacle coincides with the ray $k^2\geq \Lambda_{1}=\ds\frac{\pi^2}{4 b^2}$ which is divided by the thresholds $\Lambda_{n}=\ds\frac{\pi^2 n^2}{4b^2}, n=1,2,\cdots$; the multiplicity of this continuous spectrum is equal to $2n$ when $\Lambda_{n}<k^2<\Lambda_{n+1}$. We will consider only the case of the Neumann condition on $\Gamma$ (rigid obstacle). Note that it is known \cite{Mc} that there are no trapped modes in the interval $[0,\Lambda_{1}]$ for the soft (Dirichlet) obstacle. Moreover, we will restrict ourselves to the case of small obstacle ($\varepsilon\to+0$).

\no Problems of type (\ref{PD1}), (\ref{PD2}) were considered in numerous papers (see, e.g., \cite{Naz1}-\cite{McIL} and references therein) from the point of view of existence theorems, and asymptotic and numerical approximations. Note that the problem under consideration has many  features in common with the classical problem of the Schr\"odinger equation with a shallow potential well \cite{Sim1} and with the problem of a slightly deformed quantum waveguide \cite{Sim2, Exner}.

 The present paper is devoted to the construction of exact explicit solutions in the form of convergent series in $\varepsilon$ and $\varepsilon_1=\varepsilon\ln\varepsilon$ and subsequent analysis of the obtained formulas with the goal of extracting the leading terms of these series and formulating the conditions for the existence (or nonexistence) of trapped modes. Other geometries were studied by means of asymptotic techniques (for example, slender symmetric obstacle situated across the guide \cite{CM}). The case of a small obstacle with Dirichlet conditions on the walls was studied in \cite{SAN}, where a criterion for the existence of trapped modes on $[0,\Lambda_1)$ was established and the asymptotics of an eigenvalue in this interval was found. We show that this asymptotics is the leading form of the series mentioned above and also show its uniqueness.\\       
Further, we show that under certain symmetry conditions, there is a unique embedded eigenvalue on the interval $[\Lambda_1, \Lambda_2)$. As far as know, the existence result (which is known for an obstacle symmetric with respect to the $x$-axis, due to the decomposition trick of the domain of the corresponding differential operator; see, for example \cite{Vassiliev}) is new for an obstacle symmetric with respect to the $y$-axis, see Section\;\ref{Fc}.\\  
We obtain our results by means of explicit construction of the solutions of the corresponding system of boundary integral equations (cf. \cite{CM}) and their Fourier transforms. Similar considerations for trapped water waves can be found in \cite{PAM, PAMJ}.

\newpage

\section{Main results}\label{MR}
\setcounter{equation}{0}

In this section we formulate our main results which will be proven below. To do this, we need to introduce the following objects. Consider the exterior Neumann problem on the plane
\begin{equation}\label{Dlta}
\Delta\Psi=0~~{\rm in}~~\Omega_0,\quad \ds\frac{\partial \Psi}{\partial n}\Bigg\vert_{\Gamma}=n_2,\quad \nabla\Psi\to0~~{\rm as}~~r=\sqrt{x^2+y^2}\to\infty,
\end{equation} 
where the contour $\Gamma$ is the ``inflated'' contour $\gamma$, $\Gamma=\big\lbrace x=X(t), y=Y(t)\big\rbrace$, $\Omega_0$ is the exterior of this contour on the plane, $\ds\frac{\partial}{\partial n}$ is the derivative along the inward-looking normal to $\Gamma$, $n_2=\ds\frac{\dot{X}}{\sqrt{\dot{X}^2+\dot{Y}^2}}$ is its second component. Problem (\ref{Dlta}) describes the vertical flow of an unbounded fluid past the contour $\Gamma$. The solution of this problem is unique up to an additive constant (see, e.g., \cite{Neumann}) and
\begin{equation}\label{Psi}
\Psi={\rm const}-\mu\ds\frac{y}{r^2}-\nu\ds\frac{x}{r^2}+O\left(\ds\frac{1}{r^2}\right)~~{\rm as}~~r\to\infty.
\end{equation} 
Clearly, we can assume that const in (\ref{Psi}) is equal to $0$. The constants $\mu$ and $\nu$ are called strengths of the vertically and horizontally oriented  dipoles corresponding to (\ref{Dlta}), and,  according to \cite{Neumann},
\begin{equation}\label{mu}
\mu=\ds\frac{1}{2\pi}\Bigg(S+\int\limits_{\Gamma} n_2\Psi\;dl\Bigg),\quad\nu=\ds\frac{1}{2\pi}\int\limits_{\Gamma} n_1\Psi\;dl,
\end{equation}
where $S$ is the area bounded by $\Gamma$, $dl$ is the element of the arclength and $n_{1,2}$ are the components of the inward-looking normal to $\Gamma$. The constant $\mu$ is always positive.\\
Denote
\begin{equation}\label{a_0^ast}
a_0^{\ast}=\ds\frac{2b}{\pi}\arctan\sqrt{\ds\frac{S}{2\pi \mu}}.
\end{equation}
The following theorem describes the existence and asymptotics of discrete eigenvalues $k^2$ of (\ref{PD1}), (\ref{PD2}) in the interval $[0,\Lambda_1)$. 
\begin{teo}\label{sigma_alpha}
For sufficiently small $\varepsilon$, there exists a value of $a=a^{\ast}=a_0^{\ast}+O(\varepsilon)$ such that (\ref{PD1}), (\ref{PD2}) possesses a unique trapped mode on $[0,\Lambda_1)$ if $a>a^{\ast}$. If $a\leq a^{\ast}$, there are no trapped modes on this interval. The eigenvalue, when it exists, is analytic in $\varepsilon$ and $\varepsilon_1=\varepsilon\ln\varepsilon$, and is given by $k^2=\Lambda_1-\sigma^2$, where   
\begin{equation}\label{sigmaO(var)}
\sigma=\varepsilon^2\ds\frac{\pi^2}{4b^3}\Bigg(\pi\mu\sin^2\ds\frac{\alpha}{2}-\ds\frac{1}{2}S \cos^2\ds\frac{\alpha}{2}\Bigg)+O(\varepsilon^3\ln\varepsilon),\quad \alpha=\ds\frac{\pi a}{b}.
\end{equation}

\begin{obs}
Formula (\ref{sigmaO(var)}) was obtained in \cite{SAN}, our contribution consists in the proof of the analyticity of $\sigma$ in $\varepsilon, \varepsilon_1$ and the uniqueness of it. 
\end{obs}
\end{teo}
In the next theorem we consider the interval $\Lambda_1\leq k^2<\Lambda_2$.
\begin{teo}\label{a^{ast}+}
\begin{enumerate}
For sufficiently small $\varepsilon$, the following statements are valid:
\item[(i)] There are no eigenvalues on the interval $[\Lambda_1,\Lambda_2-\sigma_0]$ with any fixed $\sigma_0>0$. 

\item[(ii)] If $\nu\neq0$, then there are no eigenvalues on $[\Lambda_1,\Lambda_2)$.

\item[(iii)] If $\Gamma$ is symmetric with respect to the $x$-axis (this means that $a=0$ and $Y(t)$ is odd and $X(t)$ is even, see Fig.\;\ref{Figparametric_33}; in this case $\nu=0$ automatically), then there exists a unique eigenvalue on $[\Lambda_1,\Lambda_2)$ given by $k^2=\Lambda_2-\sigma^2$, where
\begin{equation}\label{sigma_O(varepsilon)}
\sigma=\varepsilon^2\;\ds\frac{\pi^3}{b^3} \mu+O(\varepsilon^3\ln\varepsilon).
\end{equation}
\item[(iv)] If $\Gamma$ is symmetric with respect to the $y$-axis (this means that $Y(t)$ is even and $X(t)$ is odd, see Fig.\;\ref{FigPARA_1}; in this case we also have $\nu=0$), then there exists a unique eigenvalue on $[\Lambda_1,\Lambda_2)$ if $a=\varepsilon a_1+O(\varepsilon^2)$, where $a_1$ is given by 
\begin{equation}\label{fa1}
a_1=\ds\frac{1}{2(2S+\pi\mu)}\ds\int\limits_{-\pi}^{\pi} \Big(Y\dot{X}-3X\dot{Y}\Big) \Big(Y-\Psi(X,Y)\Big)\;dt,
\end{equation} 
and this eigenvalue is given by (\ref{sigma_O(varepsilon)}). 
\end{enumerate}

\begin{exa}\label{X}
In the case of a slightly perturbed circle, it is possible to calculate explicitly all the objects entering formula (\ref{fa1}). For example, if $X(t)=\sin t-\ds\frac{\beta}{2} \sin 2t,~ Y(t)= -\cos t+\ds\frac{\beta}{2}\cos2t$, $\beta\ll 1$, (Fig.\;\ref{FigPARA_1}) $a_1=-\beta/12+O(\beta^2)$, see Section\;\ref{Fc}.
\end{exa}

\begin{figure}[htbp]
\centering
\includegraphics[scale=0.18]{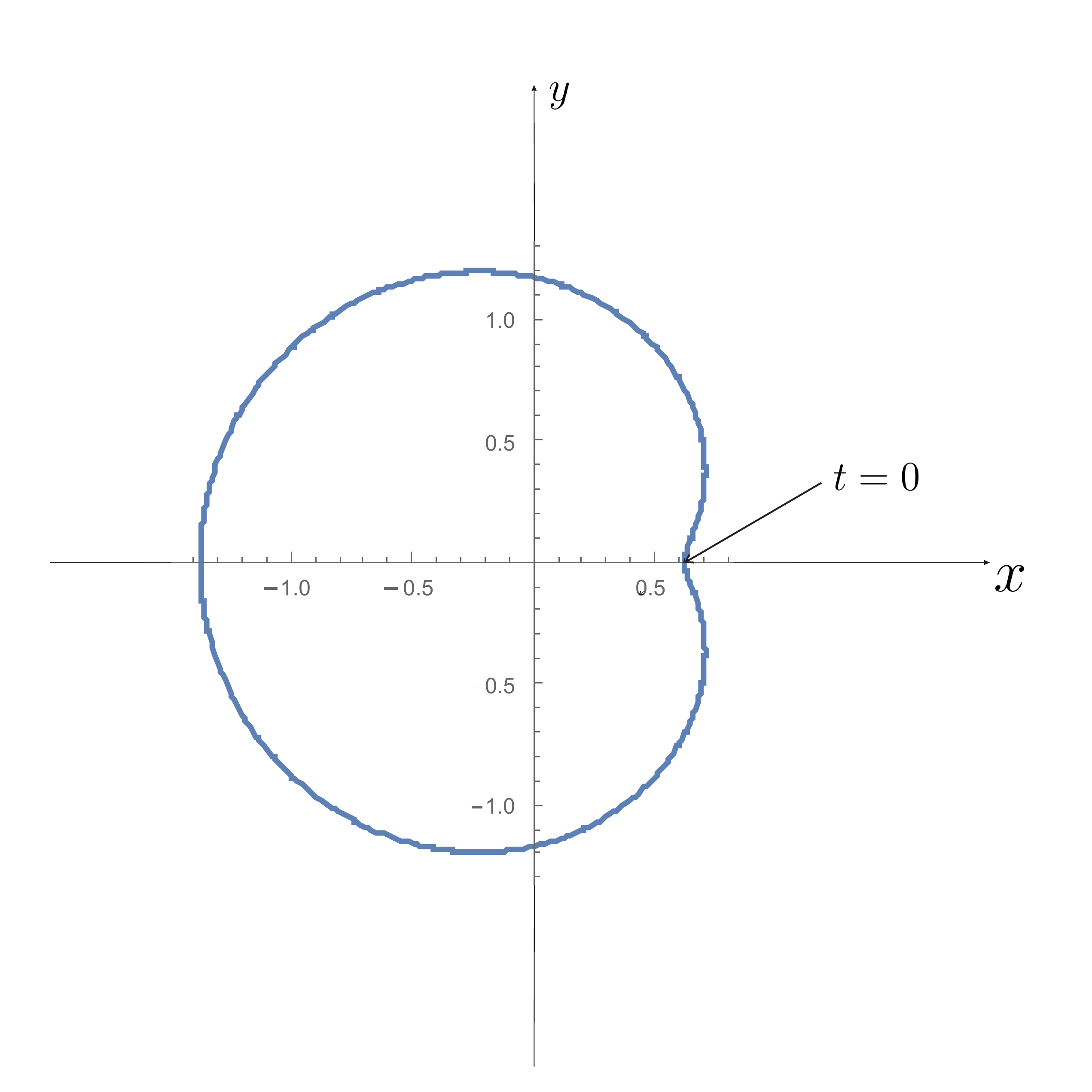}
\caption{Contour $\Gamma$ for an obstacle symmetric with respect to $x$-axis}\label{Figparametric_33}
\end{figure}

\newpage
\begin{figure}[htbp]
\centering
\includegraphics[scale=0.18]{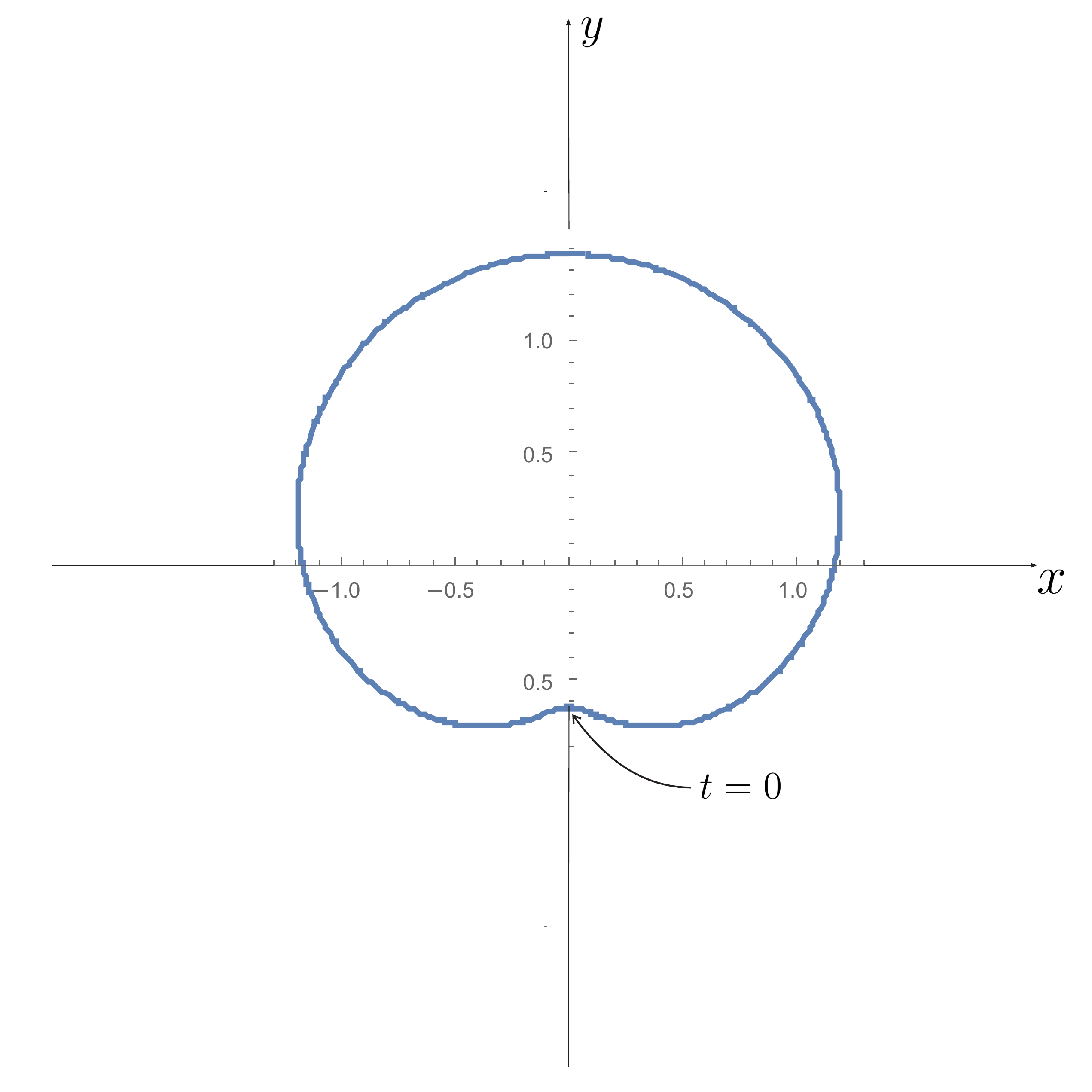}
\caption{Contour $\Gamma$ for an obstacle symmetric with respect to $y$-axis}\label{FigPARA_1}
\end{figure}
\end{teo}
\begin{obs}
The eigenvalues, when they exist, are analytic in $\varepsilon,\varepsilon_1$. The existence of the eigenvalue in $(iii)$ (even when $\varepsilon$ is not small) can be proven by means of restricting problem (\ref{PD1}), (\ref{PD2}) to solutions odd in  $y$ \cite{Vassiliev} (the restriction of problem (\ref{PD1}), (\ref{PD2}) to odd in $y$ functions removes the interval $[\Lambda_1,\Lambda_2)$ from the continuous spectrum) thus reducing the problem of embedded eigenvalues to the discrete spectrum of the restricted problem.\\Our contribution consists mainly in $(iv)$ (the restriction to odd or even in $x$ functions does not remove the interval $[\Lambda_1,\Lambda_2)$ from the continuous spectrum, thus the eigenvalue in this case is truly embedded) and the proof of uniqueness and analyticity of $\sigma$. 
\end{obs}

\begin{obs}
We note that the uniqueness of the eigenvalues is due exclusively to the smallness of the obstacle. Similarly to the one-dimensional Schr\"odinger equation with a shallow potential well, the number of eigenvalues can augment when the obstacle becomes large (just as in the case of deep potential well), as numerical results from, e.g., \cite{CM} show.
\end{obs}

\section{Boundary integral equations}\label{Ibe}
\setcounter{equation}{0}

Assuming the existence of solution to (\ref{PD1}), (\ref{PD2}) which describes trapped modes, we will reduce this problem to three integral equations for the functions $\varphi(x)$, $\psi(x)$ and $\theta(t)$, where
\begin{align*}
 \varphi(x)=\ds\frac{\partial u}{\partial n}\Big\vert_{\Gamma_{+}}=u_{y}\Big\vert_{\Gamma_{+}},\qquad \psi(x)=\ds\frac{\partial u}{\partial n}\Big\vert_{\Gamma_{-}}=-u_{y}\Big\vert_{\Gamma_{-}}, \qquad
\theta(t)=u\Big\vert_{\gamma}.
\end{align*} 
To formulate these integral equations, we will need the following objects. Introduce the Fourier transforms of $\varphi$, $\psi$ by the formulas
\begin{equation*}
\tilde{\varphi}(p)=\ds\int e^{-ipx} \varphi(x)\;dx,\qquad \tilde{\psi}(p)=\ds\int e^{-ipx} \psi(x)\;dx.
\end{equation*}
Here and everywhere below integrals without limits mean the integration over $\R$.\\
Denote ${\bf r}(t)=\Big(X(t), Y(t)\Big)$,~ ${\bf m}(t)=\Big(-\dot{Y}(t),\dot{X}(t)\Big)$.
As shown in Appendix, the Fourier transforms $\tilde{\varphi}(p)$, $\tilde{\psi}(p)$, $p\in\R$, of $\varphi$, $\psi$ and $\theta(t)$ satisfy the following system of integral equations:
\begin{equation}\label{theta+}
\theta(t)+\ds\int\limits_{-\pi}^{\pi} M(t,s) \theta(s)\; ds=\ds\int M_1(t,p)\tilde{\varphi}(p)\;dp+\ds\int M_2(t,p)\tilde{\psi}(p)\;dp
\end{equation}
where
\begin{equation}\label{M1}
M(t,s)=-\ds\frac{\varepsilon k}{2}\;\ds\frac{N'_0\Big(\varepsilon k\abs{{\bf r}(s)-{\bf r}(t)}\Big)}{\abs{{\bf r}(s)-{\bf r}(t)}}\Big({\bf r}(s)-{\bf r}(t)\Big)\cdot {\bf m}(s),
\end{equation}
$N_0(r)$ is the Neumann function, 
\begin{align}\label{KM1}
& M_1(t,p)=\ds\frac{1}{4\pi} e^{ip\varepsilon X(t)}\Bigg(\ds\frac{1}{\tau(p)} e^{-(h^{-})\tau(p)}+\ds\frac{1}{\check{\tau}(p)} e^{-(h^{-})\check{\tau}(p)}\Bigg),\qquad h^{-}=b-a-\varepsilon Y(t)\\\nonumber\\\label{KM2}
& M_2(t,p)=\ds\frac{1}{4\pi} e^{ip\varepsilon X(t)}\Bigg(\ds\frac{1}{\tau(p)} e^{-(h^{+})\tau(p)}+\ds\frac{1}{\check{\tau}(p)} e^{-(h^{+})\check{\tau}(p)}\Bigg),\qquad h^{+}=b+a+\varepsilon Y(t).
\end{align}

\begin{equation}\label{tilde{var}}
\begin{array}{lll}
\tilde{\varphi}(p)\sinh \Big(2b\tau(p)\Big)= \varepsilon \ds\int\limits_{-\pi}^{\pi} e^{-ip\varepsilon X(t)}\Bigg(ip\dot{Y}(t)\sinh \Big(h^{+} \tau(p)\Big)+\tau(p)\dot{X}(t)\cosh\Big(h^{+} \tau(p)\Big)\Bigg)\theta(t)\;dt
\end{array}
\end{equation}

\begin{equation}\label{tilde{psi}}
\begin{array}{lll}
\tilde{\psi}(p)\sinh \Big(2b\tau(p)\Big)= \varepsilon \ds\int\limits_{-\pi}^{\pi} e^{-ip\varepsilon X(t)}\Bigg(ip\dot{Y}(t)\sinh \Big(h^{-} \tau(p)\Big)-\tau(p)\dot{X}(t) \cosh\Big(h^{-}\tau(p)\Big)\Bigg)\theta(t)\;dt;
\end{array}
\end{equation}
here the functions $\tau(p)$, $\check{\tau}(p)$ are the analytic continuations of $\big(p^2-k^2\big)^{1/2}$ to the complex plane $p\in \C$ with the cuts shown in Fig.\;\ref{branch} and \ref{bra33}, respectively (see Appendix), coinciding with the arithmetical square root for $p>k$.
Note that for real $p$, $\tau(p)=\check{\tau}(p)=\sqrt{p^2-k^2}$ for $\abs{p}>k$ and  $\tau(p)=-\check{\tau}(p)=-i\sqrt{k^2-p^2}$ for $\abs{p}<k$. Note that the weak integrable singularities of the kernels $M_{1,2}(t,p)$ at the points $p=\pm k$ ($\tau=\check{\tau}=0$) do not affect the convergence of integrals in (\ref{theta+}). The kernel $M(t,s)$ is smooth.\\
Note that the functions $\theta(t)$, $\tilde{\varphi}(p)$ and $\tilde{\psi}(p)$ depend also on the spectral parameter $k$ which enters $\tau$, $\check{\tau}$ and the kernels $M, M_{1,2}$. 

\section{Discrete eigenvalue}\label{DI}
\setcounter{equation}{0}

In this section we will prove Theorem\;\ref{sigma_alpha} and construct discrete eigenvalues (if they exist) of problem (\ref{PD1}), (\ref{PD2}). This means that the spectral parameter $k^2$ satisfies $0\leq k^2<\Lambda_1$. Put
\begin{equation}\label{PD3}
k^2=\Lambda_1-\sigma^2,\quad0<\sigma\leq \pi/2b.
\end{equation}
In (\ref{PD3}) we assume that $\sigma\to 0$ as $\varepsilon\to 0$ since if $\sigma$ is bounded away from $0$, $\sigma\geq \sigma_0>0$, then, as we will see below, problem (\ref{PD1}), (\ref{PD2}) admits only the trivial solution. It is well-known $\Big($and can be easily verified by means of separation of variables in semistrips $\lbrace x>R, -b<y<b \rbrace$ and $\lbrace x<-R, -b<y<b\rbrace$ $\Big)$ that a solution of (\ref{PD1}), (\ref{PD2}) decaying as $x\to\pm\infty$ decreases exponentially together with its derivatives,
\begin{equation}\label{exp}
u=C_{\pm} e^{-\sigma|x|} \cos\ds\frac{\pi y}{2b}+O\big(e^{-\frac{\sqrt{3}}{2}\frac{\pi}{b} |x|}\big).
\end{equation}
This means, in particular, that the Fourier transforms of the Neumann data $u_{y}(x,\pm b)$ are meromorphic in the strip $\mathcal{S}=\Big\lbrace \abs{\Im p}<\sqrt{3}\pi/2b\Big\rbrace$ and have poles only at the points $p=\pm i\sigma$, since the Fourier transforms of $C_{\pm} \exp\big\lbrace -\sigma |x|\big\rbrace$ are meromorphic in this strip with the same poles $\big($of course, $\tilde{\varphi}$, $\tilde{\psi}$ can have poles or other singularities outside the strip$\big)$.\\ 
Note that it is possible to consider negative values of $\sigma$; the corresponding solutions will describe antibound states (growing at infinity). Here we will restrict ourselves only to the construction of bound states (see (\ref{exp})), that is, we consider only positive values of $\sigma$.\\

\begin{figure}[htbp]
\centering
\includegraphics[scale=0.27]{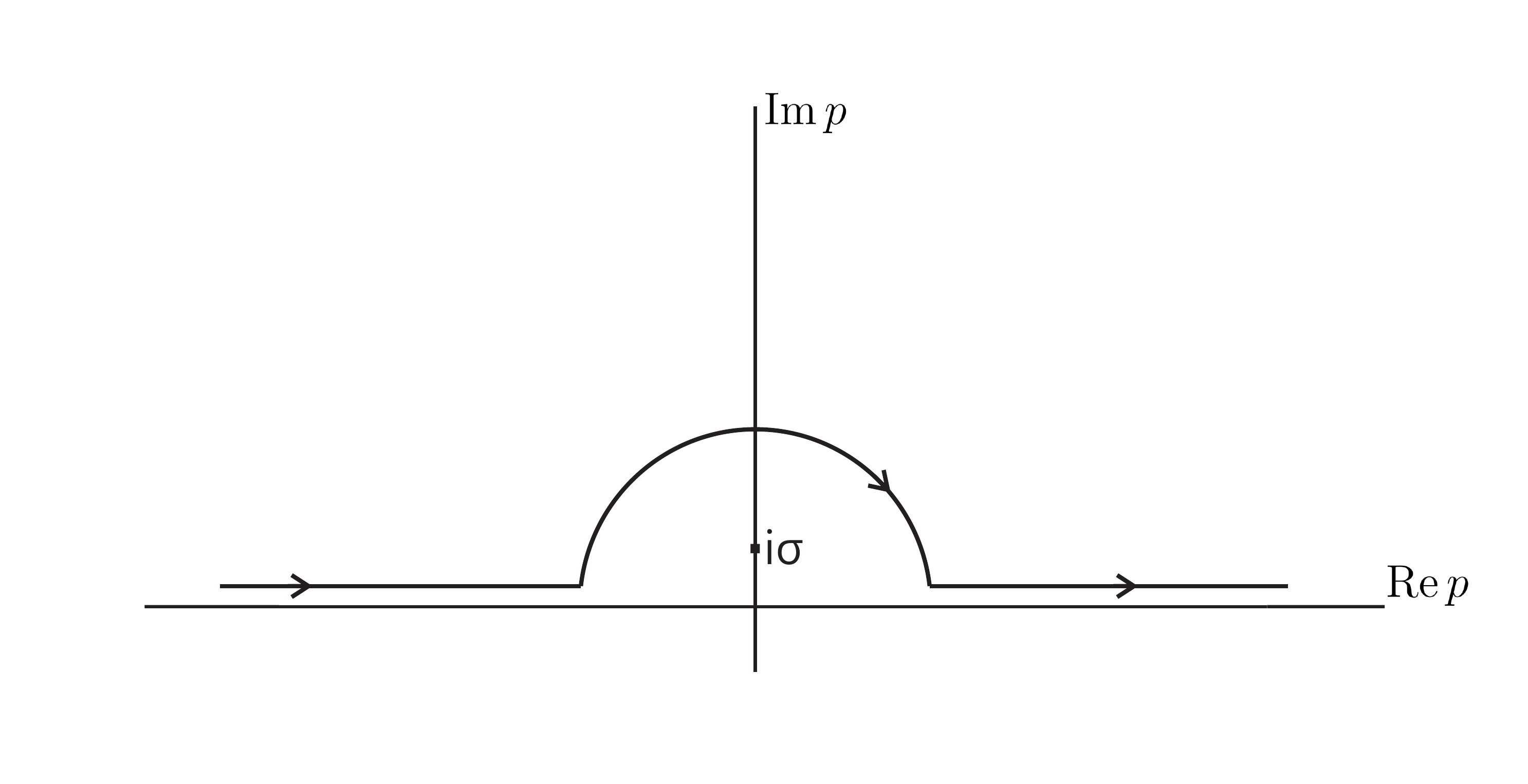}
\caption{Contour $\mathcal{C}$}\label{circum_C}
\end{figure}

Consider equations (\ref{tilde{var}}) and (\ref{tilde{psi}}). The factor $\sinh 2b\tau(p)=2\sinh b\tau\cosh b\tau$ has zeros at the points where $\sinh b\tau$ or $\cosh b\tau$ vanish.
The function $\sinh b\tau$ does not have zeros in the strip $\mathcal{S}$ apart from the points where $\tau=0$, i.e., $p=\pm k$. Indeed, $\sinh b\tau=0$ at $b\tau=\pi in$, $n=0\pm1,\cdots$ or $p^2-k^2=-\pi^2 n^2/b^2$. Since $k^2=\Lambda_1-\sigma^2$, this means that $p^2=-\ds\frac{\pi^2}{b^2}\Big(n^2-\ds\frac{1}{4}\Big)-\sigma^2$, and the last expression, for sufficiently small $\sigma$, is never positive for $n\neq0$. Therefore, equations (\ref{tilde{var}}) and (\ref{tilde{psi}}) can be divided by $\sinh \big(b\tau(p)\big)$. Thus, for example, (\ref{tilde{var}}) takes the form   
\begin{equation*}
\tilde{\varphi}(p)\cosh b\tau(p)=\ds\frac{\varepsilon}{2}\ds\int\limits_{-\pi}^{\pi} e^{-ip\varepsilon X(t)}\Bigg(i p\dot{Y}(t)\;\ds\frac{\sinh \big(h^{+} \tau(p)\big)}{\sinh \big(b\tau(p)\big)}+\tau(p)\dot{X}(t)\ds\frac{\cosh \big(h^{+}\tau(p)\big)}{\sinh \big(b\tau(p)\big)} \Bigg) \theta(t)\;dt,
\end{equation*}
and we see that the kernel in the right-hand side is analytic in $p$ (even when $\tau=0$, i.e., $n=0$ in the above formulas for zeros of $\sinh b\tau$). This follows from the fact that this kernel is in fact a function of $\tau^2$ which is obviously analytic.      
This means that $\tilde{\varphi}(p)$ is analytic in $\mathcal{S}$ apart from the points where $\cosh b\tau(p)$ vanishes. Simple zeros of $\cosh b\tau(p)$ are located at the points $\tau(p)=i\pi/2+\pi in$, $n=0,\pm 1,\cdots$, or $p^2=\ds\frac{\pi^2}{4b^2}-\ds\frac{\pi^2}{4b^2}\Big(2n+1\Big)^2-\sigma^2$, and this means that $p$ is bounded away from the real axis for $n\neq0$ and sufficiently small $\sigma$, and, for $n=0$, $p=\pm i\sigma$ and lies inside $\mathcal{S}$ for small $\sigma$. Similarly, $\tilde{\psi}(p) \cosh b\tau(p)$ is analytic in the strip $\mathcal{S}$ (see equation (\ref{tilde{psi}})). Thus we can look for $\tilde{\varphi}$ and $\tilde{\psi}$ in the form
\begin{equation}\label{phipsi}
\tilde{\varphi}=\ds\frac{A_1(p)}{\cosh (b\tau)},\qquad \tilde{\psi}=\ds\frac{A_2(p)}{\cosh (b\tau)},
\end{equation}
where $A_{1,2}(p)$ are analytic in $\mathcal{S}$.\\
Substituting  in (\ref{theta+}), (\ref{tilde{var}}) and (\ref{tilde{psi}}), dividing (\ref{tilde{var}}) and (\ref{tilde{psi}}) by $2\sinh b\tau$ we obtain 
\begin{align}\label{A1}
& A_1(p)=\varepsilon\ds\int\limits_{-\pi}^{\pi} P_3(p,t) \theta(t)\;dt
\\ \nonumber\\\label{A2}
& A_2(p)=\varepsilon\ds\int\limits_{-\pi}^{\pi}  P_4(p,t) \theta(t)\;dt
\end{align}
\begin{equation}\label{Atheta}
\theta(t)+\ds\int\limits_{-\pi}^{\pi} M(t,s)\theta(s)\;ds=\ds\int\Big(P_1(t,p) A_1(p)+P_2(t,p) A_2(p)\Big) dp,
\end{equation}
where $P_{1,2}(t,p)=M_{1,2}(t,p)\big/\cosh b\tau$,
\begin{align}\label{M6}
& P_3(p,t)=\ds\frac{1}{2} e^{-ip\varepsilon X} \Bigg(ip\dot{Y}\;\ds\frac{\sinh (h^{+} \tau)}{\sinh (b\tau)}+\dot{X}\;\ds\frac{\tau\cosh (h^{+} \tau)}{\sinh (b\tau)}\Bigg),
\\\nonumber\\\label{M7}
& P_4(p,t)=\ds\frac{1}{2} e^{-ip\varepsilon X} \Bigg(ip\dot{Y}\;\ds\frac{\sinh (h^{-} \tau)}{\sinh (b\tau)}-\dot{X}\;\ds\frac{\tau\cosh (h^{-} \tau)}{\sinh (b\tau)}\Bigg).
\end{align}
Recall that $\tilde{\varphi}$, $\tilde{\psi}$ and $\theta$ depend on $k$ and hence $A_{1,2}(p)$ and $\theta(t)$ depend also on $\sigma$.\\
These considerations lead us to the following statement. Recall that our aim is to construct trapped modes, i.e., find the values of $k^2$ (i.e., by (\ref{PD3}), the values of $\sigma$) which correspond to nontrivial solutions of (\ref{PD1}), (\ref{PD2}) describing trapped modes.
In order to solve (\ref{A1})-(\ref{Atheta}) we will use the Banach space $\mathscr{A}(S)$ of vector functions ${\bf A}=(A_1,A_2)$ analytic in $\mathcal{S}$ with the norm
\begin{equation}\label{BSA}
\norm{{\bf A}}=\max_{1,2}\Bigg(\sup_{\mathcal{S}}\Big\lbrace\abs{A_1}\exp\big(-3b\abs{p}/4\big)\Big\rbrace, \sup_{\mathcal{S}}\Big\lbrace\abs{A_2} \exp\big(-3b\abs{p}/4\big)\Big\rbrace\Bigg).
\end{equation}
\begin{lem}\label{4.1}
If problem (\ref{PD1}),(\ref{PD2}) admits a trapped mode for $0\leq k^2<\Lambda_1$, then system (\ref{A1})-(\ref{Atheta}) must possess a nontrivial solution $A_{1,2}$, $\theta, \sigma$ such that ${\bf A}(p)\in\mathscr{A}(S)$, $\theta(t)$ is continuous and $\sigma>0$. 
\end{lem}
Conversely, if (\ref{A1})-(\ref{Atheta}) possesses such a solution, then problem (\ref{PD1})-(\ref{PD2}) admits a trapped mode which can be reconstructed by means of the Green formula (\ref{GF}) since then $\tilde{\varphi}$ and $\tilde{\psi}$ given by (\ref{phipsi}) decrease exponentially  as $\abs{p}\to \infty$ in $\mathcal{S}$. 
Thus our goal consists in solving system (\ref{A1})-(\ref{Atheta}).\\
Consider equation (\ref{Atheta}). We would like to solve this equation with respect to $\theta$ and substitute its solution in (\ref{A1}), (\ref{A2}) thus reducing our system to two equations for $A_{1,2}$. Consider the kernel $M(t,s)$ (see (\ref{M1})). By \cite{Abramowitz}, $N'_0(r)$ admits the following convergent expansion valid for small $r$:   
\begin{equation*}
N'_0(r)=\ds\frac{2}{\pi}\Bigg(\ds\frac{1}{r}-\ds\frac{r}{2}\ln\ds\frac{r}{2}+\ds\frac{r}{2}\Big(\ds\frac{1}{2}-\gamma\Big)\Bigg)+O(r^3\ln r),
\end{equation*}
where the $O$-symbol is analytic in $r$ and $r\ln r$.
Hence, 
\begin{equation}\label{Mser}
M(t,s)=M^{(0)}(t,s)+\varepsilon^2\ln\varepsilon\; M^{(1)}\big(t,s,\varepsilon,\varepsilon\ln\varepsilon\big)+\varepsilon^2\; M^{(2)}\big(t,s,\varepsilon,\varepsilon\ln\varepsilon\big),
\end{equation}
where
\begin{equation}\label{M1O}
M^{(0)}=-2\ds\frac{\partial G_0\Big({\bf r}(s)-{\bf r}(t)\Big)}{\partial n} \abs{{\bf m}(s)},
\end{equation}
with $G_0(x,y)=\ds\frac{1}{2\pi} \ln r$,~$r=\sqrt{x^2+y^2}$ and $\partial/\partial n$ being the derivative along the inward-looking normal to $\Gamma$; the normal derivative is calculated at the point $\Big(X(s), Y(s)\Big)$. The kernels $M^{(1,2)}$ are smooth in $t$, $s$ (at least of class $C^1$) and analytic in $\varepsilon$, $\varepsilon_1=\varepsilon\ln\varepsilon$, and hence the integral operators with these kernels are bounded in $C[-\pi,\pi]$.\\
It is well-known (see, e.g. \cite{Petrovsky, Maz}) that the operator $1+\hat{M}^{(0)}$ is invertible in $C[-\pi,\pi]$, and hence the operator $1+\hat{M}$ also is. Denote
\begin{equation}\label{LL_0}
\hat{L}=(1+\hat{M})^{-1},\qquad \hat{L}_0=\Big(1+\hat{M}^{(0)}\Big)^{-1}.
\end{equation}
Thus (\ref{Atheta}) yields 
\begin{equation}\label{LTOA}
\theta=\hat{L} \hat{T}_0 {\bf A},
\end{equation}
where
\begin{equation}\label{TOA}
\hat{T}_0 {\bf A}=\ds\int \Big(P_1(t,p) A_1(p)+P_2(t,p) A_2(p)\Big)\;dp.
\end{equation}
Obviously, $\hat{T}_0$ acts from $\mathscr{A}$ to $C[-\pi,\pi]$.
Substituting (\ref{LTOA}) in (\ref{A1}), (\ref{A2}), we finally obtain an equation for the vector function ${\bf A}(p)$:
\begin{equation}\label{KTOA}
{\bf A}=\varepsilon \hat{K} \hat{L}\hat{T}_0\;{\bf A},
\end{equation}
where $\hat{K}$ is defined by 
\begin{equation}\label{DefK}
\hat{K}\theta=
\Big(\hat{P}_3 \theta, \hat{P}_4 \theta\Big).
\end{equation}
Recall that we are interested in a nontrivial solution of (\ref{KTOA}). Since $\cosh b\tau(p)$ does not have zeros in $\mathcal{S}$ if $\sigma$ is bounded away from $0$, $\sigma\geq \sigma_0>0$, the operator $\hat{T}_0$ is bounded uniformly in $\sigma$ for such $\sigma$. Thus equation (\ref{KTOA}) possesses only the trivial solution and hence there are no trapped modes.\\ 
On the other hand, if $\sigma$ is small ($\sigma\to 0$), the zeros $p=\pm i\sigma$ of $\cosh b\tau(p)$ are close to the real axis and the operator $\hat{T}_0$ is not bounded uniformly in $\sigma$, $\sigma_0\geq \sigma>0$.
Its ``unbounded part'' corresponds to the residue of the integrand in (\ref{TOA}) at the point $p=i\sigma$. Subtracting the principal part of the Laurent expansion in $\sigma$ of this residue from $\hat{T}_0$, we obtain a bounded (uniformly in $\sigma$) operator. Note that this principal part can be calculated explicitly and has a simple form (see (\ref{R})). (Of course, similar calculations can be performed for the other pole $p=-i\sigma$; we will use only the upper pole).\\
Let us perform these calculations.     
Consider the right-hand side of (\ref{Atheta}). By the residue theorem, we have
\begin{equation}\label{Res}
\hat{T}_0 {\bf A}=\ds\int \big(P_1 A_1+P_2 A_2\big)\;dp=\ds\int\limits_{\mathcal{C}} \big(P_1 A_1+P_2 A_2\big)\;dp+2\pi i\; \overset{}{\underset{p=i\sigma}{\Res}}\;\big(P_1 A_1+P_2 A_2\big),
\end{equation}
here $\mathcal{C}$ is the contour which circumvents the pole $p=i\sigma$ from above, see Fig.\;\ref{circum_C}.

It is easy to see, taking into account that $\tau(i\sigma)=-i\pi/2b$, $\check{\tau}(i\sigma)=i\pi/2b$ (see Section\;\ref{Ibe} and Appendix), and expanding the integrand in Taylor series with respect to $\sigma$, that  
\begin{equation*}
2\pi i\; \overset{}{\underset{p=i\sigma}{\Res}}\;\big(P_1 A_1+P_2 A_2\big)=\ds\frac{1}{\sigma}\Big(A_1(i\sigma)+A_2(i\sigma)\Big) \Big(R(t) +O(\sigma)\Big),
\end{equation*}
where
\begin{equation}\label{R}
R(t)=-\ds\frac{1}{b}\cos\ds\frac{\pi}{2b}\Big(a+\varepsilon Y(t)\Big).
\end{equation}
Hence the integral operator $\hat{T}_1$ defined by  
\begin{equation}\label{T1}
\begin{array}{lll}
\hat{T}_1{\bf A}
:= \ds\int \big(P_1 A_1+P_2 A_2\big)\;dp-\ds\frac{1}{\sigma} \Big(A_1(i\sigma)+A_2(i\sigma)\Big) R
\end{array}
\end{equation}
is bounded uniformly in $\sigma\to 0$ as an operator  from the space $\mathscr{A}(\mathcal{S})$ to $C[-\pi,\pi]$. Indeed, by (\ref{Res}), (\ref{T1}),  
\begin{equation*}
\begin{array}{lll}
\hat{T}_1 {\bf A}&=&\hat{T}_0{\bf A}-\ds\frac{1}{\sigma}\Big(A_1(i\sigma)+A_2(i\sigma)\Big) R(t)\\\\
&=&\ds\int\limits_{\mathcal{C}} \big(P_1 A_1+P_2 A_2\big)\:dp+2\pi i\; \overset{}{\underset{p=i\sigma}{\Res}}\;\big(P_1 A_1+P_2 A_2\big)-\ds\frac{1}{\sigma}  \Big(A_1(i\sigma)+A_2(i\sigma)\Big) R(t).
\end{array}
\end{equation*}
The integral operator in the right-hand side is bounded since the poles of the integrand are bounded away from $\mathcal{C}$, and the remaining part is the regular part of the Laurent expansion of the residue as $\sigma\to0$.\\
Rewrite equation (\ref{Atheta}) as 
$$
\big(1+\hat{M}\big)\theta= \hat{T}_1{\bf A}+\ds\frac{1}{\sigma}\Big(A_1(i\sigma)+A_2(i\sigma)\Big) R(t).
$$
We have
\begin{equation}\label{thN}
\theta=\hat{L} \hat{T}_1{\bf A}+\ds\frac{1}{\sigma} \Big(A_1(i\sigma)+A_2(i\sigma)\Big)\hat{L} R.
\end{equation}
Substituting in (\ref{A1}) and (\ref{A2}), we have
\begin{equation}\label{eqA}
{\bf A}=\varepsilon\hat{K} \hat{L} \hat{T}_1 {\bf A}+\ds\frac{\varepsilon}{\sigma} \Big(A_1(i\sigma)+A_2(i\sigma)\Big) \hat{K}\hat{L} R 
\end{equation}
where $\hat{K}$ is defined by (\ref{DefK}).
System (\ref{A1})-(\ref{Atheta}) implies that 
\begin{equation}\label{A1=A2}
A_1(i\sigma)=A_2(i\sigma)
\end{equation}
since the integrands of (\ref{A1}) and (\ref{A2}) coincide at $p=i\sigma$. Hence ${\bf A}$ satisfies
\begin{equation}\label{2A1}
{\bf A}-\varepsilon\hat{T} {\bf A}=2\ds\frac{\varepsilon}{\sigma} A_1(i\sigma) \hat{K}\hat{L} R,
\end{equation} 
where $\hat{T}= \hat{K} \hat{L} \hat{T}_1$ is a bounded linear operator on $\mathscr{A}(\mathcal{S})$. If $A_1(i\sigma)=0$, then the only solution of (\ref{2A1}) is trivial. Since we are interested in a nontrivial solution, we can assume that $A_1(i\sigma)\neq0$. Dividing (\ref{2A1}) by $A_1(i\sigma)$ and taking into account the homogeneity of (\ref{2A1}), we see that ${\bf A}/A_1(i\sigma)$ (that is, a solution ${\bf A}$ normalized by the condition $A_1(i\sigma)=1$)  satisfies       
\begin{equation}\label{FA}
{\bf A}-\varepsilon\hat{T} {\bf A}=2\ds\frac{\varepsilon}{\sigma} \hat{K} \hat{L} R, \quad A_1(i\sigma)=1.
\end{equation}
The solution of (\ref{FA}) is given by 
\begin{equation}\label{SA}
{\bf A}=2\ds\frac{\varepsilon}{\sigma} \Big(1-\varepsilon \hat{T}\Big)^{-1} \hat{K}\hat{L} R,
\end{equation}
and the normalization condition holds if and only if $\sigma$ satisfies the {\it secular equation} 
\begin{equation}\label{sigma}
\sigma=2\varepsilon\Bigg(\Big(1-\varepsilon\hat{T}\Big)^{-1}\hat{K} \hat{L} R\Bigg)_{1}\Bigg\vert_{p=i\sigma}:=\varepsilon F\big(\sigma, a, \varepsilon,\varepsilon_1\big),
\end{equation}
where $(\cdot)_{1}$ means the first entry of a vector.
Since (\ref{thN}) and (\ref{eqA}) are equivalent to system (\ref{A1})-(\ref{Atheta}), equations (\ref{SA}), (\ref{FA}) and (\ref{sigma}) are also equivalent to this system.
Obviously, the function $F$ in (\ref{sigma}) is analytic in all its arguments and hence equation (\ref{sigma}) possesses a unique solution $\sigma(\varepsilon)$ such that $\sigma\to 0$ as $\varepsilon\to 0$ by the Implicit Function Theorem. If this solution satisfies $\sigma(\varepsilon)>0$, then (\ref{SA}) defines, by Lemma\;\ref{4.1}, a trapped mode with $\theta$ given by (\ref{thN}).

\subsection{Solution of the secular equation}\label{SSE}

Let us prove that the solution $\sigma(\varepsilon)$ of (\ref{sigma}) is real-valued. Since $R$ is real by (\ref{R}), $\hat{L} R$ is also real by (\ref{thN}) because the kernel of $\hat{M}$ is real by (\ref{M1}). Moreover, at the point $p=i\sigma$ the function $\hat{K} \hat{L} R$ is real because $P_{3,4}(i\sigma,t)$ are real by (\ref{M6}), (\ref{M7}).\\  
Obviously, by (\ref{M6}), (\ref{M7}), the function $\hat{K}\hat{L} R$ is such that its real part is even in $p$ and its imaginary part is odd in $p$. Moreover, its value at the point $p=i\sigma$ is real for real $\sigma$. Call this property ``property EO''. We claim that the operator $\hat{T}=\hat{K} \hat{L} \hat{T}_1$ preserves this property. Indeed, it is easy to see that the operator $\hat{T}_1$ sends functions with property EO to real-valued functions,  since the real parts of $P_{1,2}(t,p)$ are even in $p$ and the imaginary parts are odd. As above, $\hat{L}$ sends real-valued functions to real-valued functions and $\hat{K}$ sends real-valued functions to EO-functions.
Hence, the operator $\hat{T}=\hat{K} \hat{L} \hat{T}_1$ preserves property EO for real $\sigma$. Thus, by induction, the right-hand side of  equation (\ref{sigma}) is purely real and its solution $\sigma(\varepsilon)$ is also real.\\
To obtain the conditions for the parameter $a$ that ensure the inequality $\sigma>0$, let us obtain the leading term of the expansion of $\sigma$ in the power series with respect to $\varepsilon$ and $\varepsilon_1$.\\ 
By (\ref{sigma}) we have 
\begin{equation}\label{ser}
\sigma=2\varepsilon\Big(\hat{K} \hat{L} R+\varepsilon \hat{T} \hat{K} \hat{L} R+\cdots\Big)_{1}\Bigg\vert_{p=i\sigma}.
\end{equation}
It turns out that $\big(\hat{K} \hat{L} R\big)\Big\vert_{p=i\sigma}=O(\varepsilon)$ and $\big(\hat{T}\hat{K}\hat{L} R\big)\Big\vert_{p=i\sigma}=O(\varepsilon)$ and so on. Indeed, for example, $\big(\hat{K} \hat{L} R\big)\Big\vert_{p=i\sigma}$ in the leading term is proportional to a linear combination of integrals of $\dot{X}$ or $\dot{Y}$, which, of course, vanish. Higher terms in $\varepsilon$ in (\ref{ser}) can be analyzed similarly. Thus it is sufficient, up to $O(\varepsilon^3)$, to consider the first summand in (\ref{ser}). Taking into account the fact that $\tau(i\sigma)=-i\pi/2d$, formulas (\ref{M6}), (\ref{M7}), and performing simple trigonometry, we obtain  
\begin{equation*}
\sigma=\ds\frac{\varepsilon}{b}\ds\int\limits_{-\pi}^{\pi} e^{\varepsilon\sigma X}\Bigg(\dot{X} \ds\frac{\pi}{2b}\sin\ds\frac{\pi}{2b}\Big(a+\varepsilon Y\Big)+\sigma\dot{Y} \cos\ds\frac{\pi}{2b}\Big(a+\varepsilon Y\Big) \Bigg)\hat{L} \cos\ds\frac{\pi}{2b}\Big(a+\varepsilon Y\Big)\; dt+O(\varepsilon^3).
\end{equation*}
In the leading term, the value of $\sigma$ in the right-hand side can be set equal to $0$. Moreover, $\hat{L}=\hat{L}_0+O(\varepsilon^2\ln\varepsilon)$ by (\ref{Mser}). Hence, denoting $\alpha=\ds\frac{\pi a}{2b}$ and expanding $\sin$ and $\cos$ in Taylor series with respect to $\varepsilon Y$, we obtain
\begin{equation*}
\sigma=\ds\frac{\varepsilon\pi}{2b^2}\ds\int\limits_{-\pi}^{\pi} \dot{X}\Big(\sin\alpha+\ds\frac{\pi\varepsilon Y}{2b}\cos\alpha\Big)+\hat{L}_0\Big(\cos\alpha-\ds\frac{\pi\varepsilon Y}{2b}\sin\alpha\Big)\; dt+O(\varepsilon^3\ln\varepsilon).
\end{equation*}
The leading term of the last integral vanishes because $\hat{L}_0 1=1/2$ (see \cite{PAMJ}) and we obtain
\begin{equation}\label{sige2}
\sigma=\ds\frac{\varepsilon^2\pi^2}{4b^3}\Bigg(\ds\frac{1}{2}\cos^2\alpha\ds\int\limits_{-\pi}^{\pi} \dot{X} Y\;dt-\sin^2\alpha \ds\int\limits_{-\pi}^{\pi} \dot{X} \hat{L}_0 Y\;dt\Bigg)+O(\varepsilon^3\ln\varepsilon).
\end{equation}
By \cite[Appendix\;3]{PAMJ}, 
\begin{equation}\label{pimu}
\ds\int\limits_{-\pi}^{\pi} \dot{X}\hat{L}_0 Y\;dt=-\pi\mu,
\end{equation}
 and, obviously, 
\begin{equation}\label{area} 
\ds\int\limits_{-\pi}^{\pi}\dot{X} Y\;dt=-S.
\end{equation} 
Substituting in (\ref{sige2}), we obtain formula (\ref{sigmaO(var)}) for any value of $\sigma$ (positive or negative). To finish the proof of Theorem\;\ref{sigma_alpha}, we need to obtain the conditions for $a$ which guarantee that $\sigma>0$. 
Clearly, the fact that the solution $\sigma(\varepsilon)$ of (\ref{sigma}) vanishes means that $F\big(0,a, \varepsilon, \varepsilon_1\big)=0$. This can be considered as an equation for $a$, and its solution $a^{\ast}$ is given by $a^{\ast}=a^{\ast}_0+O(\varepsilon)$, where $a^{\ast}$ is defined in (\ref{a_0^ast}). Since $\partial F/\partial a$ at the point $(0, a^{\ast}, 0, 0)$ is positive (as can be easily seen after an elementary calculation), we have $\sigma>0$ for $a>a^{\ast}$ and $\sigma\leq 0$ for $a\leq a^{\ast}$. By Lemma\;\ref{4.1}, we see that Theorem\;\ref{sigma_alpha} is proven.~~~$\blacksquare$

\begin{exa}
Formula (\ref{sigmaO(var)}) shows that if the expression 
\begin{equation*}
-S\cos^2\alpha +2\pi\mu\sin^2\alpha
\end{equation*}
is positive, then there exists a discrete eigenvalue of our problem given by $k^2=\ds\frac{\pi^2}{4b^2}-\sigma^2$. For example, for a circle of radius $r_0$ (i.e., $S=\pi r_0^2,~\mu=r_0^2$) this will be the case if
$$
\ds\frac{1}{2}-\ds\frac{3}{2}\cos2\alpha>0
$$
which is true if $a$, for example, is close to $b$ (the obstacle is close to the upper boundary of the guide), cf.\;\cite{SAN}. 
\end{exa}

\section{Embedded eigenvalue}\label{Eme}
\setcounter{equation}{0}

\begin{figure}[htbp]
\centering
\includegraphics[scale=0.27]{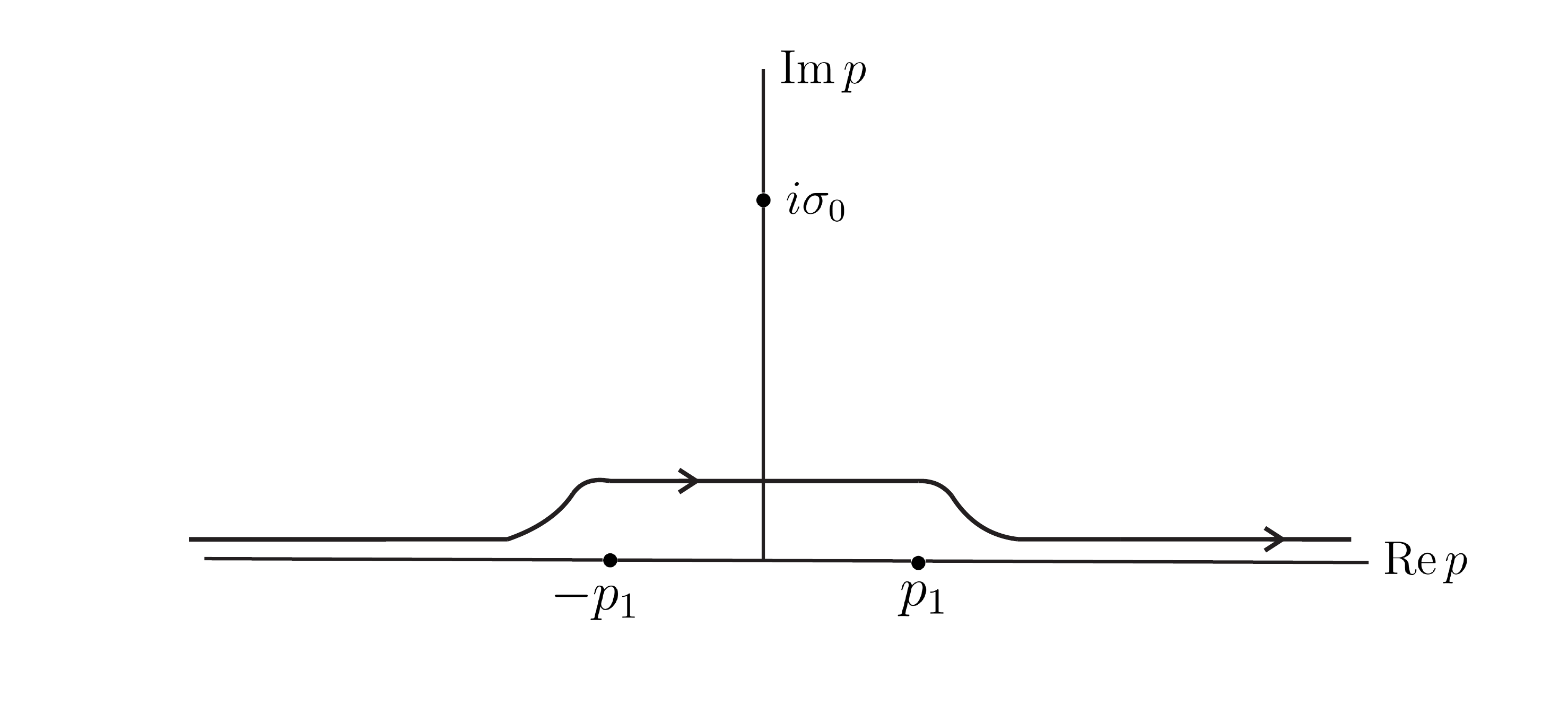}
\caption{Contour $\mathcal{C}_1$}\label{p_1_SIGMA}
\end{figure}

Let us now consider the case of eigenvalues embedded in the first segment of the continuous spectrum, $\Lambda_1\leq k^2<\Lambda_2$, that is, we will assume that $k^2=\Lambda_2-\sigma^2,  0<\sigma^2\leq\Lambda_2-\Lambda_1=3\pi^2/4b^2$.\\ 
As in Section\;\ref{DI}, a solution of (\ref{PD1}), (\ref{PD2}) decaying as $x\to\pm \infty$ decreases exponentially together with its derivatives,
\begin{equation}\label{exp1}
u=C_{\pm} e^{-\sigma\abs{x}} \sin\ds\frac{\pi y}{b}+O\big(e^{-\frac{\sqrt{5}}{2}\frac{\pi}{b}\abs{x}}\big).
\end{equation}
Of course, there also exist solutions which correspond to the plane waves 
\begin{equation*}
e^{\pm il x} \cos\ds\frac{\pi}{2b}y,\quad l=\sqrt{\ds\frac{3\pi^2}{4b^2}-\sigma^2},
\end{equation*}
but the solutions describing trapped modes cannot contain these components since they do not decrease at infinity.\\ 
Also, as above, the functions $\theta, \tilde{\varphi}, \tilde{\psi}$ must satisfy equations (\ref{theta+}), (\ref{tilde{var}}), (\ref{tilde{psi}}), and (\ref{exp1}) shows that $\tilde{\varphi}, \tilde{\psi}$ have poles at the points $p=\pm i\sigma$.\\ 
We see that $\tilde{\varphi}$ and $\tilde{\psi}$ are meromorphic in the same strip $\mathcal{S}$ as in Section\;\ref{DI} with poles at zeros of $\sinh 2b\tau=0$. This means that 
$$
2b\tau=i\pi n,\quad n=0,\pm 1,\cdots
$$
For $n=0$, $\tilde{\varphi}$ and $\tilde{\psi}$ are regular at $\tau=0$ (that is, at $p=\pm k$) since the right-hand sides of (\ref{tilde{var}}), (\ref{tilde{psi}}) vanish. For $n=\pm 1$, we have    
\begin{align*}
& \tau^2=p^2-k^2=-\ds\frac{\pi^2}{4b^2},\quad {\rm or}\quad p=\pm\sqrt{\ds\frac{3\pi^2}{4b^2}-\sigma^2}=:\pm p_1.
\end{align*}
For the values of $\sigma$ under consideration, the points $p=\pm p_1$ lie on the real line and coincide for $k^2=\Lambda_1$.\\
For $n=\pm 2$ we have
$$
p^2-k^2=-\ds\frac{\pi^2}{b^2}\quad {\rm or}\quad p^2=-\sigma^2\quad {\rm or}\quad p=\pm i\sigma.
$$
For our values of $\sigma$, the points $p=\pm i\sigma$ are bounded away from the real line for  $\sigma\geq \sigma_0>0$ and tend to $0$ as $\sigma\to0$.\\
For $|n|>2$ the zeros are bounded away from the real line, for all $\sigma$.\\
Equations (\ref{tilde{var}}) and (\ref{tilde{psi}}) show that the functions $\tilde{\varphi}\sinh(2b\tau)/\tau$ and $\tilde{\psi}\sinh(2b\tau)/\tau$ are analytic in $\mathcal{S}$ (since $\tau^{-1}\sinh 2b\tau$ is analytic in $p$ and vanishes at the poles $p=\pm i\sigma$ of $\tilde{\varphi}, \tilde{\psi}$). Thus we can look for $\tilde{\varphi}, \tilde{\psi}$ in the form
\begin{equation}\label{A12}
A_1=\ds\frac{\sinh 2b\tau}{\tau} \tilde{\varphi},\quad A_2=\ds\frac{\sinh 2b\tau}{\tau} \tilde{\psi},
\end{equation}
where $A_{1,2}$ are analytic in $\mathcal{S}$.
Since the functions $\tilde{\varphi}$, $\tilde{\psi}$ cannot have poles on the real axis, the functions $A_{1,2}(p)$ must satisfy the orthogonality conditions 
\begin{equation}\label{OC}
A_{1,2}(\pm p_1)=0.
\end{equation}
\begin{obs}
In the case $k^2=\Lambda_1$ (i.e., $\sigma^2=3\pi^2/4b^2$) we have $p_1=0$, $\sinh2b\tau(p)$ has zero of order $2$ at the origin and (\ref{OC}) should be augmented by the condition $A'_{1,2}(0)=0$.
\end{obs}
Substituting (\ref{A12}) in (\ref{theta+}), (\ref{tilde{var}}) and (\ref{tilde{psi}}), and using the fact that $\ds\frac{\tau}{\sinh 2b\tau}=\ds\frac{\check{\tau}}{\sinh 2d\check{\tau}}$, we see that $A_{1,2}$ and $\theta$ should satisfy the following three equations: 
\begin{align}\label{B_1}
& A_1=\varepsilon\ds\int\limits_{-\pi}^{\pi} Q_3(p,t)\theta(t)\;dt,
\\\nonumber\\\label{B_2}
& A_2=\varepsilon\ds\int\limits_{-\pi}^{\pi} Q_4(p,t)\theta(t)\;dt.
\end{align}

\begin{equation}\label{Btheta}
\begin{array}{lll}
\theta+\hat{M}\theta=\ds\int \Big(Q_1(t,p) A_1(p)+Q_2(t,p) A_2(p)\Big)\; dp,
\end{array}
\end{equation}
where
\begin{align*}
& Q_1(t,p)=\ds\frac{1}{4\pi} e^{ip\varepsilon X}\Bigg(\ds\frac{e^{-(h^{-} \tau)}}{\sinh 2b\tau}+\ds\frac{e^{-(h^{-}\check{\tau})}}{\sinh 2b\check{\tau}}\Bigg)\\\\
& Q_2(t,p)=\ds\frac{1}{4\pi} e^{ip\varepsilon X}\Bigg(\ds\frac{e^{-(h^{+}\tau)}}{\sinh 2b\tau}+\ds\frac{e^{-(h^{+}\check{\tau})}}{\sinh 2b\check{\tau}}\Bigg).
\end{align*}

\begin{align}\label{Q3}
& Q_3(p,t)=e^{-ip\varepsilon X}\Bigg(\ds\frac{ip\dot{Y}}{\tau} \sinh (h^{+} \tau)+\dot{X}\cosh(h^{+} \tau) \Bigg),
\\\nonumber\\\label{Q4}
& Q_4(p,t)=e^{-ip\varepsilon X}\Bigg(\ds\frac{ip\dot{Y}}{\tau} \sinh(h^{-} \tau)-\dot{X}\cosh(h^{-} \tau)\Bigg). 
\end{align}
It is easy to see that (\ref{B_1}) and (\ref{B_2}) imply that
\begin{equation}\label{EA12}
A_1(\pm p_1)=A_2(\pm p_1)
\end{equation}
since the integrands in (\ref{B_1}) and (\ref{B_2}) coincide at these points. By (\ref{OC}), the integrand in the right-hand side of (\ref{Btheta}) does not have poles on the real axis.\\
These considerations lead us to the following statement. Introduce the Banach space $\mathscr{A}(\mathcal{S})$ of vector functions ${\bf A}=(A_1,A_2)$ analytic in $\mathcal{S}$ with the norm (cf. (\ref{BSA}))
\begin{equation*}
\norm{{\bf A}}=\max_{1,2} \Bigg(\sup_{S}\Big\lbrace\abs{A_1} \exp\big(-7b\abs{p}/4\big)\Big\rbrace, \Big\lbrace\abs{A_2} \exp\big(-7b\abs{p}/4\big)\Big\rbrace\Bigg).
\end{equation*}
\begin{lem}
If problem (\ref{PD1}), (\ref{PD2}) admits a trapped mode for $\Lambda_1\leq k^2<\Lambda_2$, then system (\ref{B_1}), (\ref{B_2}), (\ref{Btheta}) must possess a nontrivial solution ${\bf A}, \theta,\sigma$ such that ${\bf A}\in \mathscr{A}(\mathcal{S})$ and satisfies (\ref{OC}), $\theta(t)$ is continuous and $\sigma>0$.  
\end{lem}
Let us continue the proof item $(i)$ of Theorem\;\ref{a^{ast}+}. We have $0<\sigma_0\leq\sigma\leq \sqrt{3}\pi/2b$. Since the integrand in (\ref{Btheta}) is analytic in $\mathcal{S}$, we can deform the integration contour to the contour $\mathcal{C}_1$ shown in Figure\;\ref{p_1_SIGMA}.  

\begin{figure}[htbp]
\centering
\includegraphics[scale=0.27]{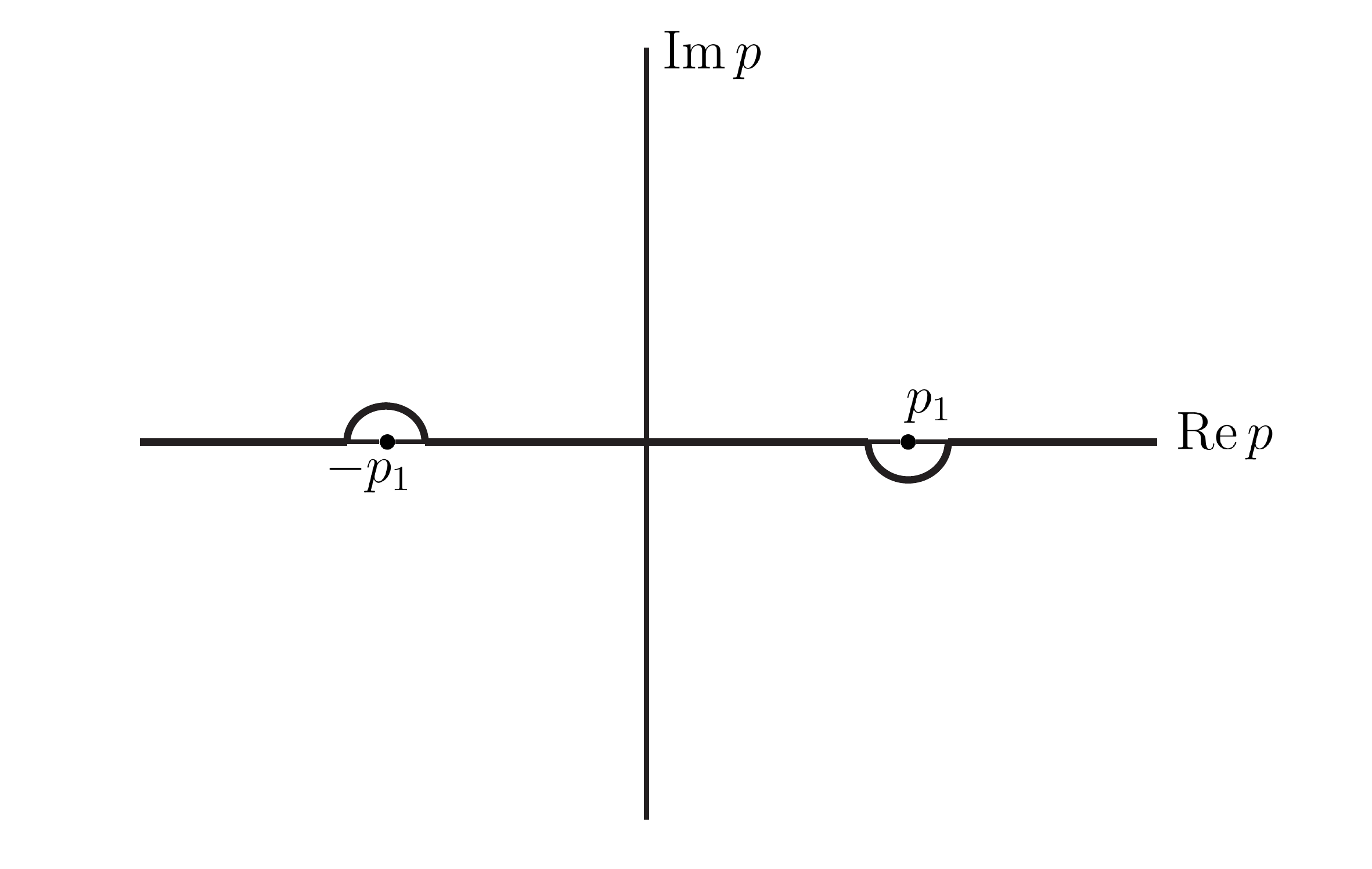}
\caption{Contour $\mathcal{C}_2$}\label{gamma}
\end{figure}

Since the poles $p=\pm p_1$ and $p=i\sigma$ of the kernel of the integral operator in (\ref{Btheta}) are bounded away from the contour $\mathcal{C}_1$, this operator is bounded from $\mathscr{A}(\mathcal{S})$ to $C[-\pi,\pi]$ and hence, as in Section\;\ref{DI} (see (\ref{LTOA}), (\ref{KTOA})), system (\ref{B_1}), (\ref{B_2}), (\ref{Btheta}) possesses only the trivial solution. Thus $(i)$ is proven.~~~~$\blacksquare$\\
Consider now items $(ii)-(iv)$. In these cases $\sigma\to 0$ as $\varepsilon \to0$. Then the poles $p=\pm p_1$ are bounded away from $p=0$ and the poles $p=\pm i\sigma$ are close to the real axis. Deform the contour of integration in (\ref{Btheta}) to the contour $\mathcal{C}_2$ shown in Figure\;\ref{gamma}.

\newpage

\begin{figure}[htbp]
\centering
\includegraphics[scale=0.27]{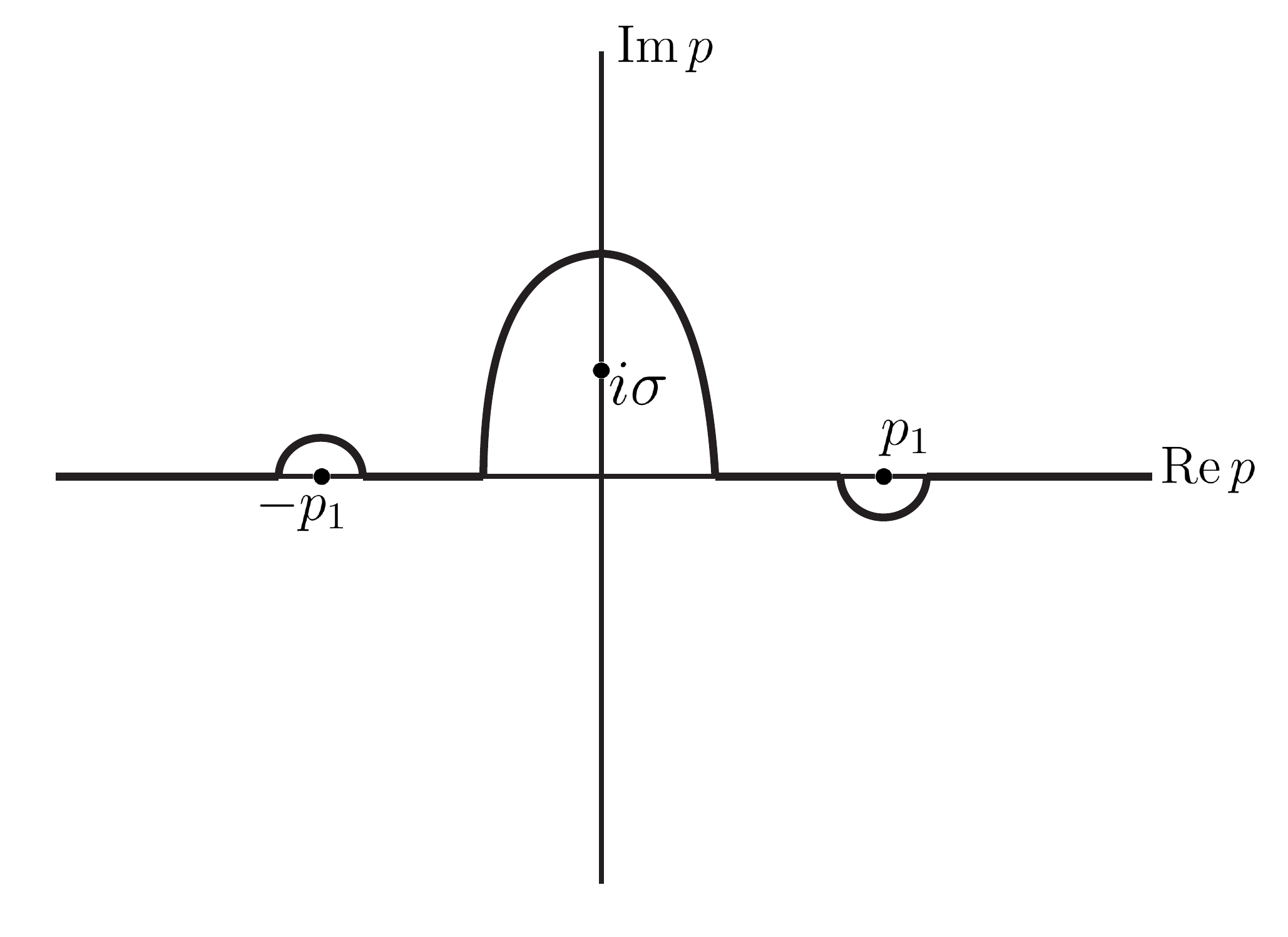}
\caption{Contour $\mathcal{C}_3$}\label{C_3}
\end{figure}

This deformation is legitimate since the integrand is analytic in $\mathcal{S}$ and allows us to avoid the poles $p=\pm p_1$ of the kernels $Q_{1,2}(t,p)$. Thus we can assume that the integral in (\ref{Btheta}) is taken over $\mathcal{C}_2$, that is, instead of (\ref{Btheta}), we consider the equation
\begin{equation}\label{Btheta2}
\theta+\hat{M}\theta=\ds\int\limits_{\mathcal{C}_2} \big(Q_1 A_1+Q_2 A_2\big) dp.
\end{equation}
Now we will apply the scheme of Section\;\ref{DI} in order to solve (\ref{B_1}), (\ref{B_2}) and (\ref{Btheta2}); later we will obtain the conditions guaranteeing that (\ref{OC}) is satisfied.\\
By the residue theorem, (\ref{Btheta}) takes the form 
\begin{equation}\label{Btheta3}
\theta+\hat{M}\theta= \ds\int\limits_{\mathcal{C}_3} \Big(Q_1(t,p) A_1(p)+Q_2(t,p) A_2(p)\Big) dp+ 2\pi i\;\overset{}{\underset{p=i\sigma}{Res}}\;\big(Q_1 A_1+Q_2 A_2\big),
\end{equation}
where the contour $\mathcal{C}_3$ is shown in Fig.\;\ref{C_3}.

\newpage

As is Section\;\ref{DI}, let us calculate the leading term of the residue in (\ref{Btheta3}). Recalling that $\tau(p)$ at the point $p=i\sigma$ has the form $\tau(i\sigma)=-i\sqrt{k^2+\sigma^2}$ where now $k^2=\ds\frac{\pi^2}{b^2}-\sigma^2$, we have $\tau(i\sigma)=-i\pi/b$ (see Section\;\ref{Ibe} and Appendix). Performing some elementary calculations as in Section\;\ref{DI}, we come to
\begin{equation*}
2\pi i\;\overset{}{\underset{p=i\sigma}{Res}}\;\big(Q_1 A_1+Q_2 A_2\big)=\ds\frac{1}{\sigma}\Big(A_1(i\sigma)-A_2(i\sigma)\Big) \Big(R(t)+ O(\sigma)\Big),
\end{equation*}
where $R(t)=\ds\frac{\pi}{2b^2} \sin\frac{\pi}{b} \Big(a+\varepsilon Y(t)\Big)$.
These calculations mean, similarly to Section\;\ref{DI}, that the operator $\hat{T}_1$ given by 
\begin{equation}\label{hat{T}1}
\hat{T}_1 {\bf A}:=\ds\int\limits_{\mathcal{C}_2} \big(Q_1 A_1+Q_2 A_2\big) dp-\ds\frac{1}{\sigma}\Big(A_1(i\sigma)-A_2(i\sigma)\Big) R(t)
\end{equation}
is bounded uniformly in $\sigma\to 0$ as an operator from $\mathscr{A}(S)$ to $C[-\pi,\pi]$. Rewriting (\ref{Btheta2}) in the form
\begin{equation}\label{thetae}
\theta+\hat{M} \theta=\hat{T}_1 A+\ds\frac{1}{\sigma}\Big(A_1(i\sigma)-A_2(i\sigma)\Big) R(t),
\end{equation}
inverting the operator $(1+\hat{M})$, and substituting $\theta$ in (\ref{B_1}), (\ref{B_2}), we finally obtain an equation for ${\bf A}$: 
$$
{\bf A}=\varepsilon\hat{K}\hat{L}\hat{T}_1{\bf A}+\ds\frac{\varepsilon}{\sigma}\Big(A_1(i\sigma)-A_2(i\sigma)\Big) \hat{K}\hat{L} R, 
$$
where
\begin{equation}\label{Ktheta}
\hat{K}\theta= \Big(\hat{Q}_3 \theta, \hat{Q}_4 \theta\Big).
\end{equation}
Since the operator $\hat{K} \hat{L}\hat{T}_1 {\bf A}$ is bounded, we have 
\begin{equation}\label{Ab}
{\bf A}=\ds\frac{\varepsilon}{\sigma} \Big(A_1(i\sigma)-A_2(i\sigma)\Big)\Big(1-\varepsilon\hat{K} \hat{L} \hat{T}_1\Big)^{-1}\; \hat{K}\hat{L} R.
\end{equation}
Similarly to Section\;\ref{DI} (see (\ref{A1=A2})), it is easy to see that
\begin{equation}\label{A1=-A2}
A_1(i\sigma)=-A_2(i\sigma)
\end{equation} 
since the integrands in (\ref{B_1}), (\ref{B_2}) coincide at $p=i\sigma$ up to a sign. Hence ${\bf A}$ satisfies
\begin{equation}\label{Afin}
{\bf A}-\varepsilon \hat{T} {\bf A}=2\ds\frac{\varepsilon}{\sigma} A_1(i\sigma) \hat{K} \hat{L} R,
\end{equation}
where $\hat{T}=\hat{K} \hat{L} \hat{T}_1$ is a bounded operator on $\mathscr{A}(\mathcal{S})$. Note that, although (\ref{Afin})coincides in form with (\ref{2A1}), this equation is rather different because in this section the operators $\hat{T}, \hat{K}$ and the function $R$ are different from those used in Section\;\ref{DI}. Similarly to (\ref{2A1}), $A_1(i\sigma)\neq0$. Dividing (\ref{Afin}) by  $A_1(i\sigma)$, we see that ${\bf A}/A_1(i\sigma)$ satisfies (cf. (\ref{FA}))
\begin{equation}\label{FAE}
{\bf A}-\varepsilon \hat{T} {\bf A}=2\ds\frac{\varepsilon}{\sigma}\hat{K}\hat{L} R,\quad A_1(i\sigma)=1.
\end{equation}
The solution of this equation is given by
\begin{equation}\label{SAE}
{\bf A}=2\ds\frac{\varepsilon}{\sigma}\big(1-\varepsilon \hat{T}\big)^{-1} \hat{K}\hat{L} R
\end{equation}
and the normalization condition $A_1(i\sigma)=1$ takes the form
\begin{equation}\label{sige}
\sigma=2\varepsilon\Bigg(\big(1-\varepsilon\hat{T}\big)^{-1} \hat{K} \hat{L} R \Bigg)_{1}\Bigg\vert_{p=i\sigma}.
\end{equation}
Equation (\ref{sige}) has a form identical to (\ref{sigma}) (although $\hat{T}_1$, $\hat{K}$ and $R$, as mentioned above, are different) with a right-hand side analytic in $\varepsilon, \varepsilon_1, \sigma$, and hence possesses a unique solution $\sigma$ such that $\sigma\to 0$ as $\varepsilon\to 0$. Moreover, by the explicit form of (\ref{sige}), $\sigma=O(\varepsilon)$. Nevertheless, we are interested only in a solution ${\bf A}$ given by (\ref{Afin}) which satisfies the orthogonality condition (\ref{OC}); as mentioned above, if (\ref{OC}) does not hold, then $\tilde{\varphi}, \tilde{\psi}$ have poles on the real axis and this is impossible if $\tilde{\varphi}, \tilde{\psi}$ correspond to a trapped mode. Thus, first, we will determine the conditions which guarantee that (\ref{OC}) is satisfied and later will calculate the corresponding values of $\sigma$. As we will see, orthogonality conditions impose certain additional conditions on the geometry of problem (\ref{PD1}), (\ref{PD2}), in contrast to the case of discrete eigenvalues.\\
Let us analyze these conditions. 
Substituting in (\ref{SAE}) $p=\pm p_1$ and dividing  by $\ds\frac{\varepsilon}{\sigma}$, we obtain that (\ref{OC}) means, by (\ref{EA12})
\begin{equation}\label{ortex}
\Big(\big(1-\varepsilon\hat{T}\big)^{-1}\;\hat{K} \hat{L} R\Big)_{1}=0\quad {\rm at}\quad p=\pm p_1.
\end{equation}
The left-hand side of this equation is analytic in $\sigma, \varepsilon, \varepsilon_1$ for small $\sigma$ and $\varepsilon$. Therefore, at least the leading term of (\ref{ortex}) must vanish.\\
In the leading term in $\varepsilon$, we have by (\ref{Ktheta})
\begin{equation}\label{ort1}
\hat{Q}_3 \hat{L} R=0\quad {\rm at}\quad p=\pm p_1.
\end{equation} 
At the point $p_1$ we have 
$$
\tau(p_1)=-i\pi/2b,\quad \check{\tau}(p_1)=i\pi/2b.
$$
Also, $p_1=p_{1}^{0}+O(\sigma^2),~p_{1}^{0}=\pi\sqrt{3}/2d$, and hence,  up to $O(\sigma^2)$ (which is in fact $O(\varepsilon^2)$), changing $p_{1}$ to $p_{1}^{0}$ and expanding $\sinh$ and $\cosh$ of a sum, we obtain for $Q_3\big(\pm p_1, t\big)$
\begin{equation}\label{Q3PM}
\begin{array}{lll}
Q_3(\pm p_1,t)= e^{\mp ip_1^0\varepsilon X}\Big(\mp\ds\frac{2b p_{1}^{0}}{\pi} \dot{Y}\; (-i)\cos\ds\frac{\pi}{2b}\big(a+\varepsilon Y\big)-\dot{X}\sin\ds\frac{\pi}{2b}\big(a+\varepsilon Y\big)\Big).
\end{array}
\end{equation}
Thus (\ref{ort1}) takes the form, up to $O(\sigma^2)=O(\varepsilon^2)$, 
\begin{equation}\label{Opm}
\begin{array}{lll}
O_{\pm}:=\hat{Q}_3\hat{L} R\Big\vert_{p=\pm p_1}&=&\ds\int\limits_{-\pi}^{\pi} e^{\mp ip_1^0\varepsilon X}\Big(\pm i\ds\frac{2b p_{1}^{0}}{\pi} \dot{Y}\; \cos\ds\frac{\pi}{2b}\big(a+\varepsilon Y\big)-\dot{X}\sin\ds\frac{\pi}{2b}\big(a+\varepsilon Y\big)\Big)\\\\
&\times&\Big(\hat{L} \sin\ds\frac{\pi}{b}\big(a+\varepsilon Y\big)\Big)dt=0.
\end{array}
\end{equation}
Expanding the exponentials and trigonometric functions in $\varepsilon$ and taking into account that $\hat{L}=\hat{L}_0+O(\varepsilon^2\ln\varepsilon)$ as above, we obtain, using (\ref{pimu}), (\ref{area}) and the facts that $\hat{L}_0 1=1/2$ and $\int_{-\pi}^{\pi} \dot{Y} \hat{L}_0 Y dt=\pi\nu$ (see \cite{PAMJ}), and recalling that $(p_{1}^{0})^{2}=3\pi^2/4b^2$, 
\begin{equation}\label{Opmc}
\begin{array}{lll}
O_{\pm}&=& \ds\frac{\varepsilon\pi}{b}\Big(2S\cos^2\ds\frac{\pi a}{2b}+\pi\mu\cos\ds\frac{\pi a}{b}\Big)\sin\ds\frac{\pi a}{2b}\pm 2i\varepsilon \ds\frac{\pi\sqrt{3}}{2b}\pi\nu\cos\ds\frac{\pi a}{2b}\cos\ds\frac{\pi a}{b}+O(\varepsilon^2\ln\varepsilon).
\end{array}
\end{equation}
Thus orthogonality condition (\ref{ortex}), in the leading term, means that the first summand in (\ref{Opmc}) is zero, i.e.,  
\begin{equation}\label{ortr}
O_{\pm}^{r}:=\Big(2S \cos^2\ds\frac{\pi a}{2b}+\pi\mu\cos\ds\frac{\pi a}{b}\Big)\sin\ds\frac{\pi a}{2b}=0
\end{equation}
and
\begin{equation}\label{orti}
O_{\pm}^{i}:=\nu \cos\ds\frac{\pi a}{2b}\cdot\cos\ds\frac{\pi a}{b}=0.
\end{equation}
Firstly, let us analyze (\ref{orti}). This equality means that 
\begin{equation*}
\cos\pi a/b=0\quad {\rm or}\quad \nu=0
\end{equation*}
since $\cos\pi a/2b\neq0$ (because $\abs{a}<b$). 
But, if $\cos\pi a/b=0$, (i.e. $a=b/2$) we have
\begin{equation*}
O_{\pm}^{r}=2S\cos^2\ds\frac{\pi a}{2b}\;\sin\ds\frac{\pi a}{2b}>0
\end{equation*}
and this contradicts (\ref{ortr}). Hence, if $\cos\pi a/d=0$, then there are no trapped modes, because (\ref{OC}) cannot be satisfied. Thus we must assume that $\nu=0$ and hence $(ii)$ from Theorem\;\ref{a^{ast}+} is proven.\\ 
Let us continue the proof of Theorem\;\ref{a^{ast}+}. We still have to satisfy (\ref{ortr}). This is possible if
\begin{equation}\label{ortr1}
2S\cos^2\ds\frac{\pi a}{2b}+\pi\mu\cos\ds\frac{\pi a}{b}=0
\end{equation}
or
\begin{equation}\label{ortr2}
\sin\ds\frac{\pi a}{2b}=0.
\end{equation}
The first possibility (\ref{ortr1}) leads to a contradiction, i.e., under this condition the solution  $\sigma(\varepsilon)$ of (\ref{sige}) satisfies $\Re \sigma<0$ and hence there are no trapped modes.  Indeed, (\ref{sige}), similarly to (\ref{ser}), implies that in fact $\sigma$ is even smaller than $O(\varepsilon)$, i.e., $\sigma=O(\varepsilon^2)$ (cf. (\ref{sige2})) and, in the leading term,    
\begin{equation*}
\sigma=2\varepsilon \hat{Q}_3\hat{L} R\Big\vert_{p=i\sigma}=2\varepsilon\hat{Q}_3\hat{L} R\Big\vert_{p=0}+O(\varepsilon\sigma). 
\end{equation*}
Similarly to Section\;\ref{SSE} (formula (\ref{sige2})), by means of simple trigonometric calculations, it is easy to obtain that    
\begin{equation}\label{sigma3}
\begin{array}{lll}
\sigma&=&\varepsilon^2\ds\frac{\pi^2}{b^3}\ds\int\limits_{-\pi}^{\pi}\Big(\ds\frac{1}{2}\sin^2\ds\frac{\pi a}{b}\;\dot{X} Y-\cos^2\ds\frac{\pi a}{b}\;\dot{X} \hat{L}_0 Y\Big) dt+O(\varepsilon^3\ln\varepsilon)\\\\
&=&  \ds\frac{\pi^2\varepsilon^2}{b^3}\Big( -\ds\frac{S}{2}\sin^2\ds\frac{\pi a}{b}+\pi\mu\cos^2\ds\frac{\pi a}{b}\Big)+O(\varepsilon^3\ln \varepsilon).
\end{array}
\end{equation}
By (\ref{ortr1}), 
\begin{equation*}
\pi\mu= -2S\cos^2\ds\frac{\pi a}{2b}\Big/\cos\ds\frac{\pi a}{b}.
\end{equation*} 
Substituting in (\ref{sigma3}), we obtain
\begin{equation*}
\sigma=-\ds\frac{\pi^2\varepsilon^2}{b^3}\;S\Big(1+3\cos\ds\frac{\pi a}{b}+4\cos^2\ds\frac{\pi a}{b}\Big)<0,
\end{equation*}
because the quadratic form in parentheses is positive definite.
Hence in the case (\ref{ortr1}) there are no trapped modes. Thus, in order for a trapped mode to exist, we have to require that (\ref{ortr2}) holds; i.e. $a=0$ in the leading term. As we have already mentioned, we have to require that  $\nu=0$. 
Thus we have proven the following statement.
\begin{prop}\label{CNC}
(Necessary condition for the existence of embedded trapped modes).\\
Embedded trapped modes for (\ref{PD1}), (\ref{PD2}) can exist only if $a=O(\varepsilon)$ and $\nu=0$.
\end{prop}
Unfortunately, we do not know in general when the condition $\nu=0$ from Proposition\;\ref{CNC} is satisfied; but we do know that if the obstacle is symmetric with respect to the $x$-axis or with respect to the $y$-axis, then $\nu=0$.\\ 
In what follows, we will assume that this symmetry condition is satisfied. Let us begin with item $(iii)$.

\subsection{Symmetry with respect to $x$-axis}

Consider the symmetry with respect to $x$-axis (Fig.\;\ref{Figparametric_33}). Then we can assume that $Y$ is odd and $X$ is even.
We will show that in this case, for $a=0$ identically, the solution ${\bf A}(p)$ given by (\ref{Ab}) is such that $A_2(p)=-A_1(p)$ and, by (\ref{EA12}), conditions (\ref{OC}) are automatically satisfied if $a=0$. We have, for $a=0$,
$$
R(t)=\ds\frac{2\pi^2}{b^2}\;\sin\ds\frac{\varepsilon \pi Y(t)}{b},
$$
\begin{align}\label{Q15}
& Q_1(t,p)=\ds\frac{1}{4\pi} e^{ip\varepsilon X}\Bigg(\ds\frac{e^{-(b-\varepsilon Y)\tau}}{\sinh 2b\tau}+\ds\frac{e^{-(b-\varepsilon Y)\check{\tau}}}{\sinh 2b\check{\tau}}\Bigg)
\\\nonumber\\\label{Q25}
& Q_2(t,p)=\ds\frac{1}{4\pi} e^{ip\varepsilon X}\Bigg(\ds\frac{e^{-(b+\varepsilon Y)\tau}}{\sinh 2b\tau}+\ds\frac{e^{-(b+\varepsilon Y)\check{\tau}}}{\sinh 2b\check{\tau}}\Bigg)
\\\nonumber\\\label{Q35}
& Q_3(p,t)=e^{-ip\varepsilon X}\Bigg(\ds\frac{ip\dot{Y}}{\tau} \sinh (b+\varepsilon Y)\tau+\dot{X}\cosh(b+\varepsilon Y)\tau\Bigg)
\\\nonumber\\\label{Q45}
& Q_4(p,t)=e^{-ip\varepsilon X}\Bigg(\ds\frac{ip\dot{Y}}{\tau} \sinh (b-\varepsilon Y)\tau-\dot{X}\cosh(b-\varepsilon Y)\tau\Bigg) 
\end{align}
Consider equation (\ref{Afin}):
\begin{equation*}
{\bf A}=\ds\frac{\varepsilon}{\sigma}\sum_{n=0}^{\infty}\Big(\varepsilon \hat{K} \hat{L} \hat{T}_1\Big)^{n} \hat{K} \hat{L} R.
\end{equation*}
Let us investigate the properties of the function
\begin{equation}\label{series}
\sum_{n=0}^{\infty}\Big(\varepsilon \hat{K} \hat{L} \hat{T}_1\Big)^{n} \hat{K} \hat{L} R.
\end{equation}
Clearly, $R(t)=\ds\frac{\pi}{2b^2}\sin\big(\varepsilon\pi Y(t)\big/b\big)$ is odd. Hence, $\hat{L} R$ is odd in $t$.
Consider the first summand in (\ref{series}), 
$$
\hat{K} \hat{L} R=
\Big(\hat{Q}_3 \hat{L} R, \hat{Q}_4  \hat{L} R\Big).
$$
It is easy to see by means of direct calculations that
\begin{equation}\label{Q3NR}
\hat{Q}_3\hat{L}R=-\hat{Q}_4\hat{L} R.
\end{equation}
Moreover, the real parts of $\hat{Q}_{3,4} \hat{L} R$ are even in $p$, and the imaginary parts are odd.\\
Let us investigate the properties of the function $\hat{T}_1 \hat{K} \hat{L} R$. 
\begin{equation}\label{T1KLR}
\hat{T}_1 \hat{K} \hat{L} R=\ds\int\limits_{\mathcal{C}_2}\Big(Q_1\hat{Q}_3\hat{L} R+Q_2\hat{Q}_4\hat{L}R\Big) dp-\ds\frac{1}{\sigma} R\Big(\hat{Q}_3\hat{L} R-\hat{Q}_4\hat{L} R\Big)\Bigg\vert_{i\sigma}.
\end{equation}
We have $Q_1(-t,p)=Q_2(t,p)$, $Q_1(t,p)=Q_2(-t,p)$ by (\ref{Q15}), (\ref{Q25}). This implies that $\hat{T}_1 \hat{K} \hat{L} R=:J(t)$ is odd in $t$. Indeed, by (\ref{Q3NR}), the first summand in (\ref{T1KLR}) has the form
\begin{equation*}
J(t)=\ds\int\limits_{\mathcal{C}_2}\big(Q_1-Q_2\big)\hat{Q}_3\hat{L} R\;dp,
\end{equation*}
and hence, because $\hat{Q}_3\hat{L} R$ is a function of $p$ alone,
\begin{equation*}
\begin{array}{lll}
J(-t)&=&\ds\int\limits_{\mathcal{C}_2}\Big(Q_1(-t,p)-Q_2(-t,p)\Big)\hat{Q}_3\hat{L} R\;dp\\\\
&=& \ds\int\limits_{\mathcal{C}_2}\Big(Q_2(t,p)-Q_1(t,p)\Big)\hat{Q}_3\hat{L} R\;dp
= -J(t).
\end{array}
\end{equation*}
In its turn, this implies that $\hat{K} \hat{L} \hat{T}_1 \hat{K} \hat{L} R$ is such that its second component is equal to the first up to multiplication by $-1$ and the real parts of the components are even in $p$ and the imaginary parts are odd. By induction, $\bf A$ given by (\ref{Afin}) has the same property, and, by (\ref{EA12}), $\bf A$ satisfies (\ref{OC}). Hence, if $a=0$ to all orders, we have a trapped mode with $\sigma$ given by (\ref{sigma3}) with $a=0$.\\   
The fact that $\sigma$ is real follows by the same argument as in Section\;2.1. Indeed, since the orthogonality condition, as we have seen, is automatically satisfied if $a=0$, the integral in the right-hand side of (\ref{Btheta}) can be taken along the real axis. Since $A_{1,2}$ possess the property EO, this integral is real and hence $\theta(t)$ is real and odd in $t$. From (\ref{B_1}) and (\ref{B_2}) it is easy to see that $A_{1,2}(i\sigma)$ are purely real and hence equation (\ref{sige}) is also purely real. Thus item $(iii)$ from Theorem \ref{a^{ast}+} is proven.  $\blacksquare$

\subsection{Symmetry with respect to $y$-axis}\label{Fc}

Consider now item $(iv)$, i.e., an obstacle symmetric with respect to $y$-axis and $a=0$ in the leading term. Thus we assume that $Y$ is even and $X$ is odd, and 
\begin{equation}\label{epsa}
a=\varepsilon a_1+O\big(\varepsilon^2\ln\varepsilon\big).
\end{equation}
Our goal is to find $a_1$ and thus determine (in the leading term) the shift of the obstacle in the vertical direction which ensures the existence of a trapped mode with $\sigma$ still given by (\ref{sigma3}) with $a=O(\varepsilon)$, i.e., 
\begin{equation*}
\sigma=\ds\frac{\pi^3\varepsilon^2}{b^3}\cdot\mu+O(\varepsilon^3\ln\varepsilon).
\end{equation*}
Let us prove first that equations (\ref{OC}) and (\ref{sige}) are purely real and hence their solutions also are. Indeed, using the symmetry in $x$, it is easy to see that ${\bf A}(p)$ is even in $p$ and $\theta(t)$ is even in $t$; obviously, $R$ is real and hence $\theta$ and ${\bf A}$ given by (\ref{thetae}) and (\ref{Afin}) are also real.\\ 
Let us calculate the first correction for the expressions from (\ref{OC}) and for the operator $\hat{L}$.\\
By (\ref{EA12}), conditions (\ref{OC}) read
\begin{align*}
& \Big(\big(1-\varepsilon \hat{T}\big)^{-1}\;\hat{K}\hat{L} R\Big)_{1}=0,\quad p=\pm p_1,
\end{align*}
that is,
\begin{equation}\label{ast}
 \Big(\hat{K} \hat{L} R\Big)_{1}+\varepsilon \Big(\hat{T} \hat{K} \hat{L} R\Big)_{1}+\cdots=0,\quad p=\pm p_1.
\end{equation}
In order to identify the leading term (which, as we will show, is of order of $O(\varepsilon^2)$), we note first that
\begin{equation}\label{6.2}
R(t)=\ds\frac{\pi}{2b^2}\sin\ds\frac{\pi}{b}\Big(a+\varepsilon Y\Big)=\varepsilon\;\ds\frac{\pi^2}{2b^3}\Big(a_1+Y\Big)+O(\varepsilon^2\ln\varepsilon).
\end{equation}
Also, $\hat{L}=\hat{L}_0+O(\varepsilon^2\ln\varepsilon)$. Since $R=O(\varepsilon)$, we can change $\hat{L}$ to $\hat{L}_0$ in (\ref{ast}) up to $O(\varepsilon^3\ln \varepsilon)$. Thus, up to the same error, the orthogonality condition reads
\begin{equation}\label{tkl_0}
\big(\hat{K} \hat{L}_0 R\big)_{1}+\varepsilon\big(\hat{T}\hat{K}\hat{L}_0 R\big)_{1}+\cdots=0,\quad p=\pm p_1.
\end{equation}  
Consider the function $\hat{K}\hat{L}_0 R$. Note that $R(t)$ is even in $t$. Hence, from the explicit form of $Q_{3,4}(p,t)$ and the fact that $\hat{L}_0$ preserves parity, it follows that $\hat{K} \hat{L}_0 R$ is even in $p$. Let us continue the calculation of the leading term of $\big(\hat{Q}_3 \hat{L}_0 R\big)\Big\vert_{p=\pm p_1}$. We have, using (\ref{6.2}), (\ref{Q3}), expanding $Q_3(p,t)$ in the Taylor series with respect to $\varepsilon$, and taking into account (\ref{epsa}) and the fact that $\hat{L}_0 1=1/2$,  
\begin{equation}\label{ortap}
\begin{array}{lll}
\Big(\hat{Q}_3\hat{L}_0 R\Big)\Big\vert_{p=\pm p_{1}}
=\varepsilon\ds\int\limits_{-\pi}^{\pi}\Big(1\mp ip_1^{0}\varepsilon X\Big)\Big(\pm i\;\ds\frac{2b p_1^0}{\pi}\dot{Y}-\ds\frac{\pi}{2b}(a_1+Y)\varepsilon \dot{X}\Big)\Big(\ds\frac{\pi a_1}{2b}+\ds\frac{\pi}{b}\hat{L}_0 Y\Big)dt+O(\varepsilon^3\ln\varepsilon).
\end{array}
\end{equation}
The coefficient at the $\varepsilon$-term in (\ref{ortap}) has the form
\begin{equation*}
\pm i\ds\int\limits_{-\pi}^{\pi} p_1^0\;\dot{Y}\Big(\ds\frac{a_1}{2}+\hat{L}_0 Y\Big) dt
\end{equation*}
and vanishes since $\dot{Y} \hat{L}_0 Y$ is odd in $t$. The coefficient of the $\varepsilon^2$-term reads
\begin{equation}\label{ortap1}
\ds\int\limits_{-\pi}^{\pi} 2(p_1^0)^2 X\dot{Y} \Big(\ds\frac{a_1}{2} +\hat{L}_0 Y\Big) dt-\ds\frac{\pi^2}{2b^2} \ds\int\limits_{-\pi}^{\pi}\big(a_1+Y\big)\dot{Y} \Big(\ds\frac{a_1}{2}+\hat{L}_0 Y\Big) dt. 
\end{equation} 
We have $(p_1^0)^2=\ds\frac{3\pi^2}{4b^2}$, and hence the orthogonality condition, in the leading term, imply that (\ref{ortap1}) vanishes, i.e., 
$$
3\ds\int\limits_{-\pi}^{\pi} X\dot{Y}\Big(\ds\frac{a_1}{2}+\hat{L}_0 Y\Big) dt-\ds\int\limits_{-\pi}^{\pi}(a_1+Y) \dot{X}\Big(\ds\frac{a_1}{2}+\hat{L}_0 Y\Big) dt=0
$$
or
\begin{equation}\label{a1}
\begin{array}{lll}
3\ds\int\limits_{-\pi}^{\pi} X\dot{Y} \hat{L}_0 Ydt-\ds\int\limits_{-\pi}^{\pi} Y\dot{X} \hat{L}_0 Ydt&=&\ds\frac{a_1}{2}\Bigg\lbrace -3\ds\int\limits_{-\pi}^{\pi} X\dot{Y} dt+\ds\int\limits_{-\pi}^{\pi} Y\dot{X} dt+2\ds\int\limits_{-\pi}^{\pi}\dot{X} \hat{L}_0 Ydt\Bigg\rbrace\\\\
&=&-a_1(2S+\pi\mu),
\end{array}
\end{equation}
by (\ref{pimu}), (\ref{area}).\\
Let us show that the next term in (\ref{tkl_0}), i.e., $\varepsilon\Big(\hat{T} \hat{K} \hat{L}_0 R\Big)_{1}\Big\vert_{\pm p_1}$ is in fact of order of $\varepsilon^3$. Indeed, as we already noted, $\hat{K} \hat{L}_0 R$ is even in $p$. From the explicit form of the operator $\hat{T}_1$ (see (\ref{hat{T}1})) it easily follows that $g(t)=\hat{T}_1\hat{K}\hat{L}_0 R$ is even in $t$.
Moreover, since $R=O(\varepsilon)$ by (\ref{6.2}), $g(t)$ is also $O(\varepsilon)$, as well as $\hat{L}_0 g$. 
In order to prove that $\Big(\hat{K}\hat{L}_0 g\Big)_{1}\Big\vert_{p=\pm p_1}=O(\varepsilon^2)$, we have to prove that 
\begin{equation}\label{Q3Neps}
\hat{Q}_3 \hat{L}_0 g\Big\vert_{p=\pm p_1}=O(\varepsilon^2). 
\end{equation}
But, by (\ref{Q3PM}),
\begin{equation}\label{Q3Ng}
\hat{Q}_3\hat{L}_0 g\Big\vert_{\pm p_1}=\ds\int\limits_{-\pi}^{\pi} e^{\mp ip_1 \varepsilon X}\Bigg(\pm \ds\frac{2b p_1 i}{\pi}\;\dot{Y}\cos\ds\frac{\pi}{2b}\Big(\varepsilon a_1+\varepsilon Y\Big)-\dot{X}\sin\ds\frac{\pi}{2b}\Big(\varepsilon a_1+\varepsilon Y\Big)\Bigg)\hat{L}_0 g\;dt+O(\varepsilon^2).
\end{equation}
From the fact that $\hat{L}_0 g$ is even, it follows that all the terms in the integrand of (\ref{Q3Ng}) which are $O(1)$, i.e., $\pm \dot{Y} \hat{L}_0 g$,   vanish because they are odd, and, since $\hat{L}_0 g=O(\varepsilon)$, we obtain (\ref{Q3Neps}). Thus, item\;$(iv)$ from Theorem\;\ref{a^{ast}+} is proven with $a_1$ given by (\ref{a1}), i.e.,  
\begin{equation}\label{6.9}
a_1=\ds\frac{1}{2S+\pi\mu}\Bigg(\ds\int\limits_{-\pi}^{\pi} Y\dot{X} \hat{L}_0 Y\;dt-3\ds\int\limits_{-\pi}^{\pi} X\dot{Y} \hat{L}_0 Y\;dt\Bigg).
\end{equation}
Using the fact that $\hat{L}_0 Y=\ds\frac{1}{2}\Big(Y-\Psi\Big\vert_{\mathcal{C}}\Big)$, where $\Psi$ is defined in (\ref{Dlta}) (see formula (7.10) in \cite{PAMJ}), we finally obtain formula (\ref{fa1}).   $\blacksquare$
\begin{exa}\label{EX}
(Calculation of $a_1$ in the case of a slightly perturbed circle, see Example\;\ref{X}).
Since the explicit values of $\mu$ and the integrals entering (\ref{6.9}) are, in general, not known, we will calculate these quantities for a slightly deformed circle shown in Fig.\;\ref{FigPARA_1},
\begin{equation}\label{XYC}
X(t)=\sin t-\ds\frac{\beta}{2}\sin 2t,\quad Y(t)=-\cos t+\ds\frac{\beta}{2}\cos2t,
\end{equation}
where $\beta>0$ is a sufficiently small parameter (in Fig.\;\ref{FigPARA_1}, for example, $\beta=0.8$). Substituting (\ref{XYC}) in (\ref{M1O}) and performing tedious but elementary trigonometric calculations, we obtain
\begin{equation*}
M^{(0)}(t,s)=\ds\frac{1}{2\pi}-\ds\frac{\beta}{2\pi} \cos s+O(\beta^2). 
\end{equation*}

Let us calculate $\hat{L}_0 Y:=f(t)$. Obviously, $f$ satisfies, by (\ref{XYC}),  
\begin{equation*}
f(t)+\ds\int\limits_{-\pi}^{\pi} M^{(0)} (t,s) f(s)\;ds=-\cos t+\ds\frac{\beta}{2}\cos2t.
\end{equation*}
By the standard perturbation theory, $f=f_0+\beta f_1+\cdots$, where $f_0(t)=-\cos t$, $f_1(t)=\ds\frac{1}{2}\Big(\cos2t-\ds\frac{1}{2}\Big)$. Obviously, $2S+\pi\mu=3\pi+O(\beta)$.  Substituting this expression and $f$ in (\ref{6.9}) and performing again elementary trigonometric calculations, we obtain $a_1=-\ds\frac{\beta}{12}+O(\beta^2)$.~~$\blacksquare$
\end{exa}

\section{Appendix. Boundary integral equations}
\begin{figure}[htbp]
\centering
\includegraphics[scale=0.27]{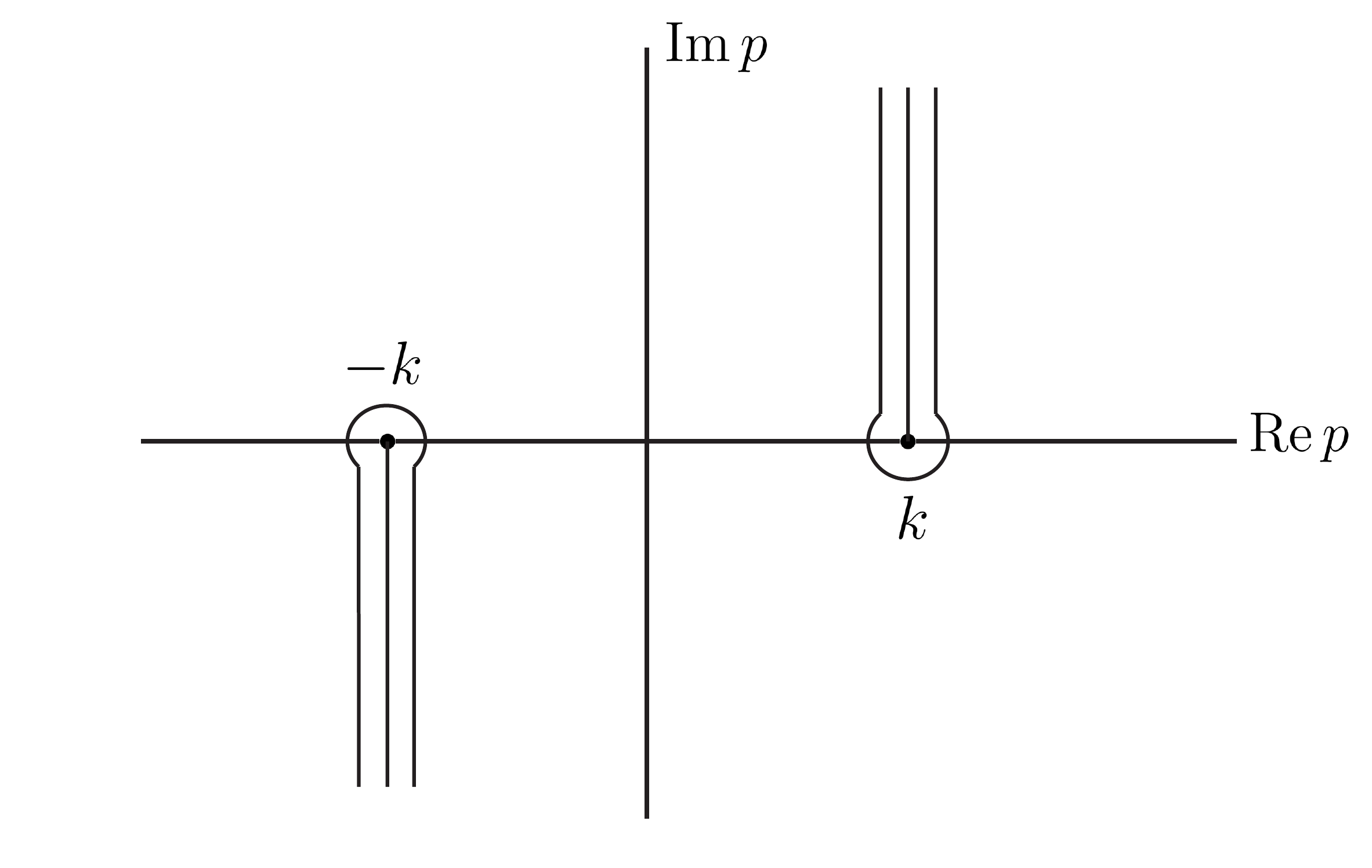}
\caption{Cuts for $\tau(p)$}\label{branch}
\end{figure}

\setcounter{equation}{0}
Here we derive system (\ref{theta+}), (\ref{tilde{var}}) and (\ref{tilde{psi}}) for the boundary values of $u$. We have, by the Green formula applied to the domain $\Omega$, 
\begin{equation}\label{GF}
u(\xi,\eta)=-\ds\int\limits_{\Gamma_{+}+\Gamma_{-}+\gamma} G\Big(x-\xi,y-\eta\Big)\ds\frac{\partial u}{\partial n} dl+\ds\int\limits_{\Gamma_{+}+\Gamma_{-}+\gamma} u\;\ds\frac{\partial G(x-\xi,y-\eta)}{\partial n} dl,\quad(\xi,y)\in\Omega,\quad(x,y)\in\partial\Omega
\end{equation}
$dl$ is the element of the arclength of $\partial\Omega$ at the point $(x,y)$, $G(x,y)$ is any fundamental solution of (\ref{PD1}) bounded at infinity, $\Delta G+k^2 G=\delta(x,y)$ and $\partial/\partial n$ is the derivative along the exterior normal to $\Omega$.\\
Denote
\begin{equation*}
\begin{array}{lll}
u_{y}\big\vert_{\Gamma_{+}}&=&\ds\frac{\partial u}{\partial n}\Big\vert_{\Gamma_{+}}=\varphi(x),\quad -u_{y}\big\vert_{\Gamma_{-}}=\ds\frac{\partial u}{\partial n}\Big\vert_{\Gamma_{-}}=\psi(x), \quad u\Big\vert_{\gamma}=\theta(t).
\end{array}
\end{equation*}
Passing to the limits  $\xi,\eta\to\Gamma_{\pm}, \gamma$ in (\ref{GF}), we obtain by (\ref{PD2}) the following three integral equations for the functions $\varphi$, $\psi$ and $\theta$ taking into account that $u(x,\pm b)=0$:\\ 
For the limit on $\Gamma_{+}\Big({\rm i.e.}~\xi\in\R,~\eta\to b\Big)$:
\begin{equation}\label{fephi}
\begin{array}{lll}
u(\xi, b)=0&=& -\ds\int G(x-\xi,0)\;\varphi(x) dx-\ds\int G(x-\xi,-2b)\;\psi(x) dx+\\\\
&+& \ds\int\limits_{\gamma}\theta(t)\;\ds\frac{\partial G}{\partial n}\Big(\varepsilon X(t)-\xi, -h^{-}\Big) dl,
\end{array}
\end{equation}
$dl$ is the arc element of $\gamma$ at the point $\varepsilon X(t), a+\varepsilon Y(t)$.\\
For the limit on $\Gamma_{-}\Big({\rm i.e.}~\xi\in\R,~\eta\to -b\Big)$:
\begin{equation}\label{sephi}
\begin{array}{lll}
u(\xi-b)=0&=&-\ds\int G(x-\xi,2b)\;\varphi(x) dx-\ds\int G(x-\xi,0)\;\psi(x) dx+\\\\
&+&\ds\int\limits_{\gamma}\theta(t)\;\ds\frac{\partial G}{\partial n}\Big(\varepsilon X(t)-\xi, h^{+}\Big) dl.
\end{array}
\end{equation}
For the limit on $\gamma \Big({\rm i.e.}~\xi\to\varepsilon X(t),~\eta\to a+\varepsilon Y(t)\Big)$, using the jump conditions, we have  
\begin{equation}\label{tephi}
\begin{array}{lll}
\theta(t)&=& -\ds\int G(x-\varepsilon X, h^{+})\;\varphi(x) dx- \ds\int G(x-\varepsilon X, -h^{-})\;\psi(x) dx\\\\
&+&\ds\int\limits_{-\pi}^{\pi}\theta(s)\ds\frac{\partial G}{\partial n}\Bigg(\varepsilon X(s)-\varepsilon X(t), \varepsilon Y(s)-\varepsilon Y(t)\Bigg)\varepsilon \sqrt{\dot{X}^2(s)+\dot{Y}^2(s)} ds+\ds\int\limits_{\gamma}\theta\;\ds\frac{\partial G}{\partial n} dl +\ds\frac{1}{2}\theta(t).
\end{array}
\end{equation}
Recall that, as in (\ref{KM1}), (\ref{KM2}),   $h^{\pm}=d\pm a\pm \varepsilon Y(t)$.
In equations (\ref{fephi}) and (\ref{sephi}), we use the fundamental solution of the form $G(x,y)=\ds\frac{1}{4i}\;H_{0}^{(1)}(k r)$, and in equation (\ref{tephi}) we use the fundamental solution of the form  $G(x,y)=\ds\frac{1}{4}\;N_{0}(k r)$; here $r=\sqrt{x^2+y^2}$ and $H_0^{(1)}$ is the Hankel function of the first kind and $N_0$ is the Neumann function.\\ 
Note that this procedure is justified since (\ref{fephi})-(\ref{tephi}) must be valid for any fundamental solution. The choice of the Neumann function in (\ref{tephi}) is due to the fact that $N_0$ is real-valued, and this simplifies the proof of the real-valuedness of the parameters $\sigma$ and $a$ in Theorems\; \ref{sigma_alpha}, \ref{a^{ast}+}.\\ 
Introduce the notation 
\begin{equation}\label{mr}
{\bf r}(t)=\Big(X(t), Y(t)\Big), \quad {\bf m}(t)=\Big(-\dot{Y}(t), \dot{X}(t)\Big),
\end{equation}
\begin{align*}
& {\bf r}_{\pm}(\xi,t)=\Big(\varepsilon X(t)-\xi, a+\varepsilon Y(t)\mp b\Big),\quad l_{\pm}(\xi,t)=|{\bf r}_{\pm}(\xi,t)|,\\\\
&  l(x,\xi)=\sqrt{(x-\xi)^2+4b^2}.
\end{align*}
Then equations (\ref{fephi})-(\ref{tephi}) read as (we omit the superindex $(1)$ of $H_0^{(1)}$ for brevity) 
\begin{equation}\label{fe}
\begin{array}{lll}
0&=&-\ds\frac{1}{4i}\ds\int H_{0}\Big(k|x-\xi|\Big) \varphi(x) dx-\ds\frac{1}{4i}\ds\int H_{0}\Big(k l(x,\xi)\Big)\psi(x) dx+\\\\
&+& \ds\frac{\varepsilon k}{4i}\ds\int\limits_{-\pi}^{\pi} \ds\frac{ H'_{0}\Big(k l_{+}(\xi,t)\Big)}{l_{+}(\xi,t)}\; {\bf r}_{+}(\xi,t)\cdot {\bf m}(t)\;\theta(t)\;dt,\quad\xi\in\R,
\end{array}
\end{equation}

\begin{equation}\label{se}
\begin{array}{lll}
0&=&-\ds\frac{1}{4i}\ds\int H_{0}\Big(k l(x,\xi)\Big)\varphi(x) dx-\ds\frac{1}{4i}\ds\int H_{0}\Big(k|x-\xi|\Big) \psi(x) dx+\\\\
&+& \ds\frac{\varepsilon k}{4i}\ds\int\limits_{-\pi}^{\pi} \ds\frac{ H'_{0}\Big(k l_{-}(\xi,t)\Big)}{l_{-}(\xi,t)}\; {\bf r}_{-}(\xi,t)\cdot {\bf m}(t)\;\theta(t)\;dt, \quad\xi\in\R,
\end{array}
\end{equation}

\begin{equation}\label{te}
\begin{array}{lll}
\ds\frac{1}{2} \theta(t)&=& -\ds\frac{1}{4}\ds\int N_{0} \Big(k l_{+}(x,t)\Big)\varphi(x) dx-\ds\frac{1}{4}\ds\int N_{0} \Big(k l_{-}(x,t)\Big)\psi(x) dx+\\\\
&+& \ds\frac{\varepsilon k}{4}\ds\int\limits_{-\pi}^{\pi}\ds\frac{N'_0\Big(\varepsilon k|{\bf r}(s)-{\bf r}(t)|\Big)}{|{\bf r}(s)-{\bf r}(t)|}\; \Big({\bf r}(s)-{\bf r}(t)\Big)\cdot{\bf m}(s)\;\theta(s)\;ds,\quad t\in[-\pi,\pi].
\end{array}
\end{equation}
Rewrite equations (\ref{fe})-(\ref{te}) in the form (we put the $\varepsilon$-terms in (\ref{fe}) and (\ref{se}) in the RHS and the terms containing $\theta$ in (\ref{te}) in the LHS)
\begin{equation}\label{fe1}
\begin{array}{lll}
\ds\frac{1}{4i}\ds\int H_0\Big(k |x-\xi|\Big)\varphi(x) dx+\ds\frac{1}{4i}\ds\int H_0\Big(k l(x,\xi)\Big)\psi(x) dx\\\\
=\ds\frac{\varepsilon k}{4i}\ds\int\limits_{-\pi}^{\pi} \theta(t)  \ds\frac{H'_0\Big(k l_{+}(\xi,t)\Big)}{l_{+}(\xi,t)}\;{\bf r}_{+}(\xi,t)\cdot {\bf m}(t) dt, \quad\xi\in\R
\end{array}
\end{equation}

\begin{equation}\label{se1}
\begin{array}{lll}
\ds\frac{1}{4i}\ds\int H_0\Big(k l(x,\xi)\Big)\varphi(x) dx+\ds\frac{1}{4i}\ds\int H_0\Big(k |x-\xi|\Big)\psi(x) dx\\\\
=\ds\frac{\varepsilon k}{4i}\ds\int\limits_{-\pi}^{\pi} \theta(t) \ds\frac{H'_0\Big(k l_{-}(\xi,t)\Big)}{l_{-}(\xi,t)}\;{\bf r}_{-}(\xi,t)\cdot{\bf m}(t) dt,\quad\xi\in\R
\end{array}
\end{equation}

\begin{equation*}
\begin{array}{lll}
\theta(t)+\ds\int\limits_{-\pi}^{\pi} M(t,s,\varepsilon)\theta(s) ds=\\\\
 =-\ds\frac{1}{2}\ds\int N_{0}\Big(k l_{+}(x,t)\Big) \varphi(x) dx
- \ds\frac{1}{2}\ds\int N_{0}\Big(k l_{-}(x,t)\Big) \psi(x) dx, \quad t\in[-\pi,\pi]
\end{array}
\end{equation*}

\begin{equation}\label{mM1}
M(t,s,\varepsilon)=-\ds\frac{\varepsilon k}{2}\ds\frac{N'_0 \Big(\varepsilon k\big|{\bf r}(s)-{\bf r}(t)\big|\Big)}{\big|{\bf r}(s)-{\bf r}(t)\big|}\big({\bf r}(s)-{\bf r}(t)\big)\cdot{\bf m}(s).
\end{equation}
Let us now convert equations (\ref{fe1})-(\ref{mM1}) into equation for $\tilde{\varphi}, \tilde{\psi}$ and $\theta$, taking the Fourier transform of (\ref{fe1}) and (\ref{se1}), and expressing the right-hand side of (\ref{mM1}) through $\tilde{\varphi}, \tilde{\psi}$.\\
To this end, we will need the Fourier transform of the functions $H_0^{(1,2)}\big(k\sqrt{x^2+y^2}\big)$ with respect to $x$.
By \cite{Gradshtein-Ryzhik} (formulas 6.677.3 and 6.677.4) 
\begin{align*}
& \ds\int\limits_{0}^{\infty} J_0\big(\alpha r\big)\cos\beta x\;dx=
\begin{cases}
\ds\frac{\cos\big(y\sqrt{\alpha^2-\beta^2}\big)}{\sqrt{\alpha^2-\beta^2}},\quad 0<\beta<\alpha\\\\
0,\quad\quad\quad\quad\quad\quad\quad\quad 0<\alpha<\beta
\end{cases}\\\\
& \ds\int\limits_{0}^{\infty} N_0\big(\alpha r\big)\cos\beta x\;dx=
\begin{cases}
\ds\frac{\sin\big(y\sqrt{\alpha^2-\beta^2}\big)}{\sqrt{\alpha^2-\beta^2}},\quad \quad\quad ~0<\beta<\alpha\\\\
-\ds\frac{1}{\sqrt{\beta^2-\alpha^2}} e^{-y\sqrt{\beta^2-\alpha^2}},\quad 0<\alpha<\beta.
\end{cases}\\\\
& {\rm Here}~r=\sqrt{x^2+y^2},~~~y\geq 0.
\end{align*}
Therefore, since $H_0^{(1)}(x)=J_0(x)+iN_0(x)$,
\begin{equation*}
\begin{array}{lll}
\ds\int\limits_{-\infty}^{\infty} H^{(1)}_0\big(k r\big) e^{-ipx}\;dx=2\ds\int\limits_{0}^{\infty} H^{(1)}_0\big(k r\big)\cos px\;dx\\\\
=2\ds\int\limits_{0}^{\infty} J_0\big(k r\big)\cos px\;dx+2i\ds\int\limits_{0}^{\infty} N_0\big(k r\big)\cos px\;dx\\\\
=\begin{cases}
\ds\frac{2}{\sqrt{k^2-p^2}}\; e^{i y\sqrt{k^2-p^2}},\quad\quad |p|<k\\\\
-\ds\frac{2i}{\sqrt{p^2-k^2}}\; e^{-y\sqrt{p^2-k^2}},\quad |p|>k.
\end{cases}
\end{array}
\end{equation*}
We will be interested in the continuation of these formulas to the complex values of $p$.\\
Introduce the function $\tau(p)=\sqrt{p^2-k^2}$ as the branch of $\sqrt{p^2-k^2}$ on $\C$ with cuts as in  Fig.\;\ref{branch} and such that $\tau$ coincides with the arithmetical root for $p>k$.

\begin{figure}[htbp]
\centering
\includegraphics[scale=0.27]{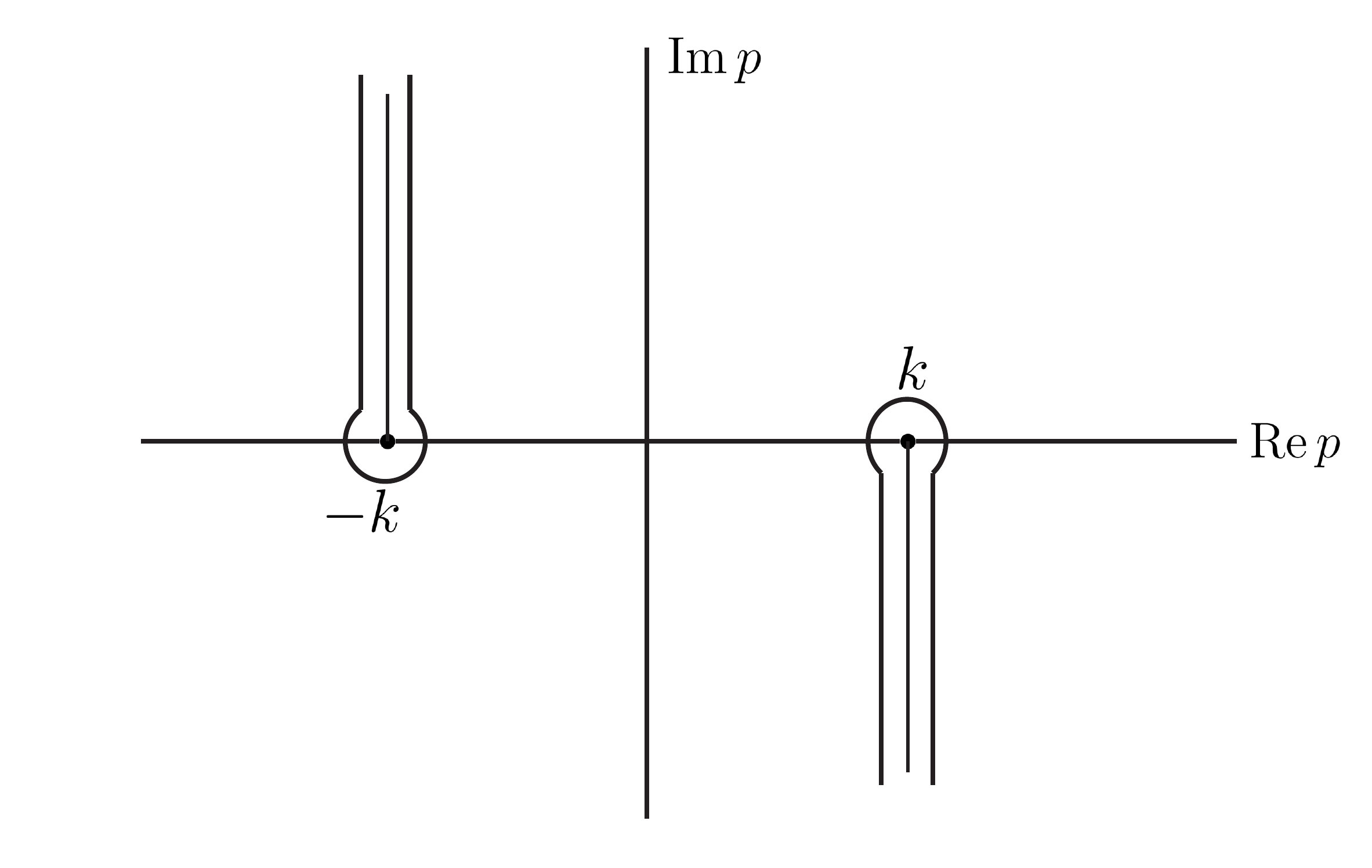}
\caption{Cuts for $\check{\tau}(p)$}\label{bra33}
\end{figure}

Then $\ds\frac{2}{\sqrt{k^2-p^2}}\;e^{iy\sqrt{k^2-p^2}}$ (with the arithmetical root) can be written as 
\begin{equation}\label{kp}
\ds\frac{2}{\sqrt{k^2-p^2}}\;e^{iy\sqrt{k^2-p^2}}=-\ds\frac{2i}{\tau}\;e^{-y\tau},\quad  |p|<k.
\end{equation}
Indeed, $\tau(p)$ for $|p|<k$ is equal to $-i\sqrt{k^2-p^2}$ and (\ref{kp}) follows. Hence
\begin{equation}\label{ft1}
\ds\int H_0\big(k r\big) e^{-ipx}\;dx=-\ds\frac{2i}{\tau}\;e^{-y\tau},\quad p\in\R,\quad p\neq\pm k.
\end{equation}
This implies that
\begin{align}\label{ft2}
& \ds\int H'_0\big(k r\big)\ds\frac{k y e^{-ipx}}{r}\;dx=-\ds\frac{2}{i} e^{-y\tau},\quad y>0
\\\nonumber\\\label{ft3}
& \ds\int H'_0\big(k r\big)\ds\frac{k x e^{-ipx}}{r}\;dx=\ds\frac{2p}{\tau} e^{-y\tau},\quad y>0.
\end{align}
In the same manner, for $H^{(2)}_0(kr)=J_0(kr)-iN_0(kr)$ we have
\begin{equation*}
\begin{array}{lll}
\ds\int H_0^{(2)} \big(k r\big) e^{-ipx}\;dx&=&2\ds\int\limits_{0}^{\infty} H_0^{(2)} \big(k r\big) \cos px\;dx\\\\
&=&\begin{cases}
\ds\frac{2}{\sqrt{k^2-p^2}}\;e^{-iy\sqrt{k^2-p^2}},\quad |p|<k\\\\
\ds\frac{2i}{\sqrt{p^2-k^2}}\;e^{-y\sqrt{p^2-k^2}},\quad |p|>k
\end{cases}\\\\
&=&\ds\frac{2i}{\check{\tau}(p)}\; e^{-y\;\check{\tau}(p)},
\end{array}
\end{equation*}
where $\check{\tau}(p)$ is the branch of the function $\sqrt{p^2-k^2}$ with cuts as in the Fig.\;\ref{bra33} and coinciding with the arithmetical root for $p>k$.

Similarly, for the Fourier transform of $N_0(kr)$ we have
\begin{equation}\label{FTNpHI}
\begin{array}{lll}
\ds\int N_0\big(k r\big) e^{-ipx}\;dx&=&\ds\frac{1}{2i}\Bigg(\ds\int H_0^{(1)}\big(k r\big) e^{-ipx}\;dx-\ds\int H_0^{(2)}\big(k r\big) e^{-ipx}\;dx\Bigg)\\\\
&=&-\Bigg(\ds\frac{1}{\tau}\;e^{-y\tau}+\ds\frac{1}{\check{\tau}}\;e^{-y\check{\tau}}\Bigg),\quad p\in\R,
\end{array}
\end{equation}
and
\begin{align}\label{6.6}
 \ds\int N'_0 \big(k r\big)\ds\frac{kx}{r}\;e^{-ipx}\;dx=(-1) ip \Bigg(\ds\frac{1}{\tau}\;e^{-y\tau}+\ds\frac{1}{\check{\tau}}\;e^{-y\check{\tau}}\Bigg)
\end{align}

\begin{equation}\label{6.7}
\ds\int N'_0 \big(k r\big)\ds\frac{ky}{r}\;e^{-ipx}\;dx=e^{-y\tau}+e^{-y\check{\tau}}. 
\end{equation}
We have, using (\ref{ft1})-(\ref{6.7}), 
\begin{equation}\label{fe2}
\ds\frac{1}{4i}\;\ds\frac{2}{i\tau}\; \tilde{\varphi}+\ds\frac{1}{4i}\;\ds\frac{2}{i\tau}\;e^{-2b\tau }\; \tilde{\psi}=\varepsilon\ds\int\limits_{-\pi}^{\pi} M_3(p,t)\theta(t)\;dt,
\end{equation}
where
\begin{equation*}
\begin{array}{lll}
M_3(p,t)&:=&\ds\frac{1}{4i}F_{x\to p}\Bigg[\ds\frac{k H'_0\Big(k l_{+}(x,t)\Big)}{l_{+}(x,t)}\;{\bf r}_{+}(x,t)\cdot {\bf m}(t)\Bigg]\\\\
&=& -\ds\frac{1}{2}\Big(\ds\frac{ip\dot{Y}}{\tau}+\dot{X}\Big) e^{-h^{-}\tau-ip\varepsilon X}.
\end{array}
\end{equation*}

\begin{equation}\label{se2}
\ds\frac{1}{4i}\;\ds\frac{2}{i\tau}\;e^{-2b\tau} \tilde{\varphi}+\ds\frac{1}{4i}\;\ds\frac{2}{i\tau}\; \tilde{\psi}=\varepsilon\ds\int\limits_{-\pi}^{\pi} M_4(p,t)\theta(t)\;dt,
\end{equation}
where
\begin{equation*}
\begin{array}{lll}
M_4(p,t)&:=&\ds\frac{1}{4i}F_{x\to p}\Bigg[\ds\frac{k H'_0\Big(k l_{-}(x,t)\Big)}{l_{-}(x,t)} {\bf r}_{-}(x,t)\cdot{\bf m}(t)\Bigg]\\\\
&=& -\ds\frac{1}{2}\Big(\ds\frac{ip\dot{Y}}{\tau}-\dot{X}\Big) e^{- h^{+}\tau-ip\varepsilon X}.
\end{array}
\end{equation*}

\medskip
Now let us rewrite equation (\ref{mM1}) in terms of $\tilde{\varphi}$ and $\tilde{\psi}$. We have
\begin{equation}\label{theta++}
\theta+\ds\int\limits_{-\pi}^{\pi} M\;\theta ds=\ds\int M_1(t,p)\tilde{\varphi}(p)\;dp+\ds\int M_2(t,p)\tilde{\psi}(p)\;dp
\end{equation}
where
\begin{align*}
& M_1(t,p)=-\ds\frac{1}{2}\ds\int N_0\Big(k l_{+}(x,t)\Big) e^{ipx}\; dx\\\\
& M_2(t,p)=-\ds\frac{1}{2}\ds\int N_0\Big(k l_{-}(x,t)\Big) e^{ipx}\; dx.
\end{align*}
Using (\ref{FTNpHI}), we obtain (\ref{KM1}), (\ref{KM2}).
Multiply (\ref{fe2}) and (\ref{se2}) by $-2\tau e^{b\tau}$:
$$
\begin{cases}
\tilde{\varphi} e^{b\tau}+\tilde{\psi} e^{-b\tau}=\varepsilon \ds\int\limits_{-\pi}^{\pi}\big(ip\dot{Y}+\tau\dot{X}\big) e^{(a+\varepsilon Y)\tau-ip\varepsilon X}\theta(t)\;dt\\\\
\tilde{\varphi} e^{-b\tau}+\tilde{\psi} e^{b\tau}=\varepsilon \ds\int\limits_{-\pi}^{\pi}\big(ip\dot{Y}-\tau\dot{X}\big) e^{-(a+\varepsilon Y)\tau-ip\varepsilon X}\theta(t)\;dt.
\end{cases}
$$
Solving this system for $\tilde{\varphi}$, $\tilde{\psi}$, we obtain (\ref{tilde{var}}), (\ref{tilde{psi}}). Taking into account (\ref{theta++}), this completes the derivation of system (\ref{theta+}), (\ref{tilde{var}}) and (\ref{tilde{psi}}) for $\tilde{\varphi}$, $\tilde{\psi}$ and $\theta$. 

\section{Acknowledgements}

PZ and AM are grateful to CONAHCYT-M\'exico and CIC-UMSNH, JEDM is grateful to CONAHCYT-M\'exico, and MIRR is grateful to Vicerrector\'ia de la Investigaci\'on de la UMNG (project INV-CIAS 3751) for partial financial support.

\end{document}